\documentclass[preprint, twocolumn]{aastex63}
\usepackage{float, bm, graphicx, amsmath, morefloats}
\usepackage[caption=false]{subfig}
\shorttitle{SN~2018erx}
\shortauthors{Das et al.}

\usepackage{float,graphicx,amsmath,tabularx,booktabs,natbib,threeparttable}
\usepackage{float, bm, graphicx, amsmath, morefloats}
\usepackage[caption=false]{subfig}
\usepackage{float,graphicx,amsmath,tabularx,booktabs,natbib,threeparttable}
\usepackage{subfig}
\usepackage[para]{footmisc}
\usepackage{tablefootnote}

\usepackage{graphicx}
\usepackage[caption=false]{subfig}
\usepackage{lipsum} 
\usepackage{placeins}

\newcommand{\lpipe}{\textsc{lpipe}}

\usepackage{comment}
\usepackage{graphicx}	
\usepackage{amsmath}	

\definecolor{dark-red}{rgb}{0.4,0.15,0.15}
\definecolor{dark-blue}{rgb}{0.15,0.15,0.4}
\definecolor{medium-blue}{rgb}{0,0,0.5}
\hypersetup{
    colorlinks, linkcolor={dark-red},
    citecolor={dark-blue}, urlcolor={medium-blue}
}

\newcommand{\beqa}{\begin{eqnarray}} 
\newcommand{\eeqa}{\end{eqnarray}}

\newcommand{\bsub}{\begin{subequations}}
\newcommand{\esub}{\end{subequations}}
\newcommand{\beal}{\begin{align}}
\newcommand{\ealn}{\end{align}}

\newcommand{\Msun}{{\ensuremath{\mathrm{M}_{\odot}}}}
\newcommand{\Rsun}{{\ensuremath{\mathrm{R}_{\odot}}}}

\graphicspath{{./}{figures/}}

\begin{document}

\title{\vspace{0.5cm} 
SN~2018erx: A fast-evolving, dust-reddened Type Icn supernova with broad C II emission lines

}

\author[0000-0001-8372-997X]{Kaustav~K.~Das}\affiliation{Cahill Center for Astrophysics, California Institute of Technology, MC 249-17, 
1200 E California Boulevard, Pasadena, CA, 91125, USA}

\author{Anjasha Gangopadhyay}
\affiliation{The Oskar Klein Centre, Department of Astronomy, Stockholm University, AlbaNova, SE-10691 Stockholm, Sweden}

\author[0000-0002-5619-4938]{Mansi~M.~Kasliwal}
\affiliation{Cahill Center for Astrophysics, 
California Institute of Technology, MC 249-17, 
1200 E California Boulevard, Pasadena, CA, 91125, USA}

\author[0000-0003-1546-6615]{Jesper Sollerman}
\affiliation{The Oskar Klein Centre, Department of Astronomy, Stockholm University, AlbaNova, SE-10691 Stockholm, Sweden}


\author[0000-0001-8472-1996]{Daniel A.~Perley}
\affiliation{Astrophysics Research Institute, Liverpool John Moores University, IC2,  Liverpool L3 5RF, UK}

\author[0000-0001-6797-1889]{Steve Schulze}
\affiliation{Center for Interdisciplinary Exploration and Research in Astrophysics (CIERA), 1800 Sherman Ave., Evanston, IL 60201, USA}

\author[0000-0002-4223-103X]{Christoffer Fremling}
\affil{Caltech Optical Observatories, California Institute of Technology, Pasadena, CA 91125, USA}

\author[0000-0002-8262-2924]{Michael W. Coughlin}
\affiliation{School of Physics and Astronomy, University of Minnesota, Minneapolis, MN 55414}

\author{Kishalay De}
\affiliation{Department of Astronomy and Columbia Astrophysics Laboratory, Columbia University, New York, NY 10027, USA}
\affiliation{Center for Computational Astrophysics, Flatiron Institute, 162 5th Ave., New York, NY 10010, USA}

\author{Avishay Gal-Yam}
\affiliation{Department of Particle Physics and Astrophysics, Weizmann Institute of Science, 234 Herzl St, 76100 Rehovot, Israel}

\author{Ariel Goobar}
\affiliation{The Oskar Klein Centre, Department of Astronomy, Stockholm University, AlbaNova, SE-10691 Stockholm, Sweden}

\author[0000-0002-3168-0139]{Matthew Graham}
\affiliation{Cahill Center for Astrophysics, 
California Institute of Technology, MC 249-17, 
1200 E California Boulevard, Pasadena, CA, 91125, USA}

\author[0000-0002-8532-9395]{Frank J. Masci}
\affiliation{IPAC, California Institute of Technology, 1200 E. California
             Blvd, Pasadena, CA 91125, USA}

\author{Takashi J. Moriya}
\affiliation{Astronomical Science Program, Graduate Institute for Advanced Studies, SOKENDAI, 2-21-1 Osawa, Mitaka, Tokyo 181-8588, Japan}
\affiliation{National Astronomical Observatory of Japan, National Institutes of Natural Sciences, 2-21-1 Osawa, Mitaka, Tokyo 181-8588, Japan}
\affiliation{School of Physics and Astronomy, Monash University, Clayton, VIC 3800, Australia}

\author{Josiah Purdum}
\affiliation{Caltech Optical Observatories, California Institute of Technology, Pasadena, CA 91125, USA}

\author[0000-0003-4725-4481]{Sam Rose}
\affiliation{Cahill Center for Astrophysics, 
California Institute of Technology, MC 249-17, 
1200 E California Boulevard, Pasadena, CA, 91125, USA}

\author[0000-0001-7648-4142]{Ben Rusholme}
\affiliation{IPAC, California Institute of Technology, 1200 E. California
             Blvd, Pasadena, CA 91125, USA}

\author[0000-0003-2700-1030]{Nikhil Sarin}
\affiliation{The Oskar Klein Centre, Department of Astronomy, Stockholm University, AlbaNova, SE-10691 Stockholm, Sweden}

\author[0000-0001-7062-9726]{Roger Smith}
\affiliation{Caltech Optical Observatories, California Institute of Technology, Pasadena, CA  91125, USA}

\author[0000-0002-6347-3089]{Daichi Tsuna}
\affiliation{Center for Astrophysics $|$ Harvard \& Smithsonian, 60 Garden St, Cambridge, MA 02138, USA}





\begin{abstract}
We present the discovery and characterization of SN~2018erx (ZTF18abkmbpy), a fast-evolving, unusually red, interacting stripped-envelope supernova. Spectroscopically, SN~2018erx exhibits prominent broad \ion{C}{2} emission with characteristic widths of $\sim\!3800$~km~s$^{-1}$, consistent with interaction with a carbon-rich circumstellar medium (CSM) and a Type~Icn core-collapse SN classification. 
Photometrically, SN~2018erx evolves exceptionally rapidly, rising from half-maximum to peak in 2.1~d and declining back in 3.1~d, placing it among the fastest-evolving core-collapse supernovae observed. Semi-analytical CSM-interaction modeling favors a compact, shell-like CSM with $M_{\rm CSM}\approx0.3\,M_\odot$, an inner radius of $R_0\approx0.7$~AU, and a very low ejecta mass of $M_{\rm ej}\approx0.11\,M_\odot$. The radioactive yield is likewise small, with a $^{56}$Ni mass $\lesssim(3$--$5)\times10^{-3}\,M_\odot$, placing SN~2018erx at the low end of the H-poor stripped-envelope SN distribution. At +29~d after peak, we detect a near-infrared excess consistent with pre-existing local circumstellar dust and infer a hot-dust mass of $M_{\rm d}\sim10^{-6}$--$10^{-5}\,M_\odot$. Taken together, the rapid evolution, strong local reddening, carbon-rich emission, and evidence for pre-existing dust point to a multi-component circumstellar environment consisting of a dense inner interaction region produced by enhanced pre-SN mass loss ($\dot{M}\sim10^{-3}$--$10^{-4}\,M_\odot\,{\rm yr^{-1}}$) and an outer dusty layer at $R\gtrsim4\times10^{16}$~cm, produced by an earlier mass-loss episode roughly $10$--$200$~yr before core collapse. These properties favor an ultra-stripped core-collapse explosion of a low-mass He star in a binary system that underwent late-time Case~BC mass transfer or Si-flash-induced mass loss. Fallback-modified Wolf--Rayet collapse or a merger-driven mass-loss episode remain plausible alternatives. These observations provide rare insight into the mass-loss history of stripped-envelope SNe, indicate that SN~2018erx-like events may be efficient sites of dust formation, and suggest that dust-enshrouded explosions of this kind may be systematically underrepresented in optical surveys.

\end{abstract}

\section{Introduction} \label{sec:intro}

Wide-field, high-cadence time-domain surveys have transformed the study of explosive transients by enabling the systematic discovery of rare and fast-evolving events. Facilities such as the Zwicky Transient Facility (ZTF; \citealt{Bellm2019,Graham2019, Dekany20, Masci2019}) and the Asteroid Terrestrial-impact Last Alert System (ATLAS; \citealt{Tonry2018,Smith2020}) now deliver large samples of core-collapse supernovae (SNe) with well-sampled photometric coverage and spectroscopy. One of the most important outcomes of this observational shift has been the growing realization that the final months to years of massive-star evolution can be extraordinarily dynamic. In many events, the explosion is accompanied by clear evidence for substantial circumstellar material (CSM) in the immediate environment, implying intense pre-SN mass loss. When the SN ejecta collide with this dense CSM, kinetic energy is efficiently converted into radiation, reshaping the light curve and imprinting narrow or intermediate-width emission features on the spectrum.

Among the diverse interaction-powered transients, the interacting stripped-envelope (SE) SNe are particularly valuable because they probe mass loss after hydrogen removal, when the progenitor has already transitioned to a compact helium star or Wolf--Rayet (WR)-like object \citep{Smith2014}. Classical SESNe (Types IIb/Ib/Ic) are understood as explosions of stars that have lost most or all of their hydrogen-rich envelopes through wind or binary stripping. Their progenitor may also have recently expelled dense material that remains sufficiently close to the star to contribute to the luminosity. The best-known subclass in this category is the Type~Ibn SNe, defined by the presence of narrow helium emission lines (often with P-Cygni profiles) and weak or absent hydrogen features, signaling interaction with He-rich CSM \citep[e.g.,][]{Pastorello2008,Hosseinzadeh2017,Farias2025}. The newly established Type~Icn SN class extends this picture further. These events show narrow or intermediate-width carbon and nitrogen lines (commonly high-ionization species at early phases), while lacking strong H and He features, pointing to interaction with H/He-poor and likely C/O-rich material. Only six confirmed SNe~Icn have been published to date—SN~2019hgp, SN~2019jc, SN~2021csp, SN~2021ckj,  SN~2022ann  and SN~2023xgo \citep{Galyam2022,Pellegrino2022,perley2022,Nagao2023,Davis2023, Gangopadhyay2025}. Even within this small sample, the class exhibits substantial diversity in luminosity, timescale, and line morphology \citep{Pellegrino2022,Nagao2023,Davis2023}, suggesting that more than one progenitor pathway may contribute to the observed phenomenology. 

The progenitor systems and mass-loss histories that give rise to SNe~Ibn and SNe~Icn remain uncertain. A natural interpretation is that both SNe~Ibn and SNe~Icn are linked to WR-like progenitors, with apparent compositional differences reflecting the surface chemistry of the star at the time of mass loss (e.g., WN-like for He-rich material versus WC/WO-like for C/O-rich material), together with variations in the geometry, timing, and density of the pre-SN outflow \citep{Galyam2022,PerleyIcn}. Lightcurve modeling of many normal SESNe often implies ejecta masses of only a few solar masses, frequently below expectations for single-star WR progenitors, motivating binary stripping as a dominant channel for producing SE explosions \citep[e.g.,][]{Drout2013,Lyman2016,Prentice2019, Taddia2018, Barbarino21_type_ic, Dessart2022}. In the most rapidly evolving events, the characteristic diffusion times can be so short that the inferred ejecta masses fall well below $\sim 1~M_{\odot}$, reminiscent of ``ultra-stripped'' explosions, which are binary-stripped He-stars with $M_\mathrm{ej} < 1\,$\Msun\ in compact binaries \citep[e.g.,][]{Tauris2013,Tauris2015}. In this regime, even modest amounts of nearby CSM can dramatically influence the observed lightcurve, and interaction-powered models have been shown to reproduce the fast and luminous peaks of several rapidly evolving transients \citep{Pellegrino2022, Wu2022b,Moriya2025}. Whether interacting SESNe preferentially arise from classical WR stars with enhanced pre-SN mass loss and fallback to a compact object, from low-mass helium stars in close binaries undergoing extreme late-stage stripping or from a mixture of these possibilities remains an open question. Another interesting progenitor channel was recently suggested by \cite{Wu2024, Moriya2026}, which suggests that white dwarf mergers can also lead to the formation of He and C+O dense CSM in SNe~Ibn/Icn. 

Several SNe~Ibn/Icn have also exhibited strong evidence for dust formation or for pre-existing dust in the immediate environment. The prototypical SN~2006jc showed a steepening optical light-curve decline accompanied by the emergence of a near-infrared (NIR) excess, a combination widely interpreted as due to rapid dust condensation in the post-shock cool dense shell \citep[e.g.,][]{Anupama2009}.
Similar dust-related behaviour has been suggested in other members of the Type Ibn class, including SN~2010al and the slowly evolving OGLE-2012-SN-006 \citep{Gan2021}. More recently, \citet{Yamanaka2025} reported a pronounced and persistent NIR excess in SN~2023xgo (classified as a Type~Ibn/Icn event) lasting from $\sim$15 to 100~days, which they interpret as a NIR echo from pre-existing circumstellar carbon dust. These cases motivate to consider local dust as an important ingredient when interpreting unusually red interacting SESNe such as SN~2018erx.

In this paper, we present SN~2018erx (ZTF18abkmbpy), a fast-evolving interacting SESN discovered by the Zwicky Transient Facility (ZTF). Its lightcurve shows an exceptionally rapid rise and decline, placing it among the fastest H-poor SNe and overlapping with the regime of rapidly evolving stripped-envelope transients \citep[e.g.,][]{Drout2014, De2018c, Yao2020, Das2024}. Prior to extinction correction, SN~2018erx is unusually red and optically faint for an interacting stripped-envelope event. Its spectra reveal prominent broad C\,\textsc{ii} emission lines, linking it to the emerging Type~Icn class, for which only six events are currently known. A late-time NIR excess further suggests the presence of pre-existing local circumstellar dust.

Taken together, these properties make SN~2018erx a rare but important event for understanding how rapid evolution, carbon-rich CSM interaction, and dust obscuration shape the observed stripped-envelope SN population. The combination of a dense inner CSM and a more extended dusty CSM component provides a direct probe of multi-epoch mass loss in heavily stripped progenitors. The carbon-rich environment also provides favorable conditions for efficient dust formation and growth. These results suggest that some fast, low-luminosity, low-$^{56}$Ni explosions may be underrepresented in current optical samples, especially if they are additionally dimmed by local dust extinction. SN~2018erx therefore provides rare insight into stripped progenitor systems, their recent mass-loss history, and possible observational biases affecting the discovery of interacting stripped-envelope explosions.


The paper is organized as follows. In Section~\ref{sec:data}, we describe the discovery of SN~2018erx and summarize the photometric and spectroscopic follow-up observations. Section~\ref{sec:extinction} presents the extinction analysis. Section~\ref{sec:observables} details the observed photometric and spectroscopic properties, including comparisons with literature transients. In Section~\ref{sec:analysis}, our analysis and modeling. In Section~\ref{sec:discussion}, we discuss the progenitor and evolutionary pathways. We summarize our conclusions in Section~\ref{sec:conclusion}.

\section{Discovery and follow-up observations}
\label{sec:data}
\subsection{Discovery and host-galaxy}
\label{sec:discovery_2018erx}

SN~2018erx was discovered with the ZTF camera \citep{Dekany20} on the 48-inch Samuel Oschin Telescope at Palomar Observatory. The source was first detected at $\alpha=17^{\rm h}05^{\rm m}14.284^{\rm s}$, $\delta=+56^{\circ}13'00.11''$ (J2000) on 2018 August 3 at 07:52:19~UT and reported to the Transient Name Server\footnote{https://www.wis-tns.org/object/2018erx/discovery-cert} as SN~2018erx \citep{Fremling2018b}. At discovery, the transient had an AB magnitude of $r_{\rm ZTF}=18.7$~mag. A pre-discovery ZTF image obtained on 2018 July 31 shows no source at the transient position down to a $5\sigma$ limiting magnitude of $r_{\rm ZTF}>20.6$~mag, implying that the rise occurred within the preceding few days.

SN~2018erx is projected on the galaxy CGCG~277$-$025 (NED position
$\alpha=17^{\rm h}05^{\rm m}13.9592^{\rm s}$,
$\delta=+56^{\circ}13'10.462''$, J2000). We adopt a host-galaxy redshift
of $z=0.02938$, measured from the narrow nebular emission lines in our
Keck/LRIS spectrum at the SN position. Assuming a flat
$\Lambda$CDM cosmology \citep{Planck2020} with $H_0=67.7~{\rm km~s^{-1}~Mpc^{-1}}$, $\Omega_{\rm m}=0.31$, this corresponds to a luminosity distance of
$D_L=133.1$~Mpc and a distance modulus of $\mu=35.62$~mag. We ignore explicit corrections for peculiar motions for galaxies at this distance. Assuming a characteristic peculiar velocity of $\sim300$~km~s$^{-1}$, the resulting uncertainty in the distance modulus is $\lesssim0.1$~mag. 
The transient is offset by $10.7''$ from CGCG~277$-$025, corresponding to a projected separation of $\sim6.5$~kpc at this distance.


We note that CGCG~277$-$025 also hosted the Type~Ib SN~2018cfh, discovered earlier in 2018 only a few weeks before SN~2018erx. The two events therefore constitute a near-contemporaneous ``sibling'' SN pair in the same host galaxy \citep[e.g.,][]{Biswas2022,Dhawan2025}. The same galaxy has since produced another core-collapse event, the Type~IIb SN~2025abku \citep{Das2025abku}.

\begin{figure*}
    \centering
    \includegraphics[width=14cm]{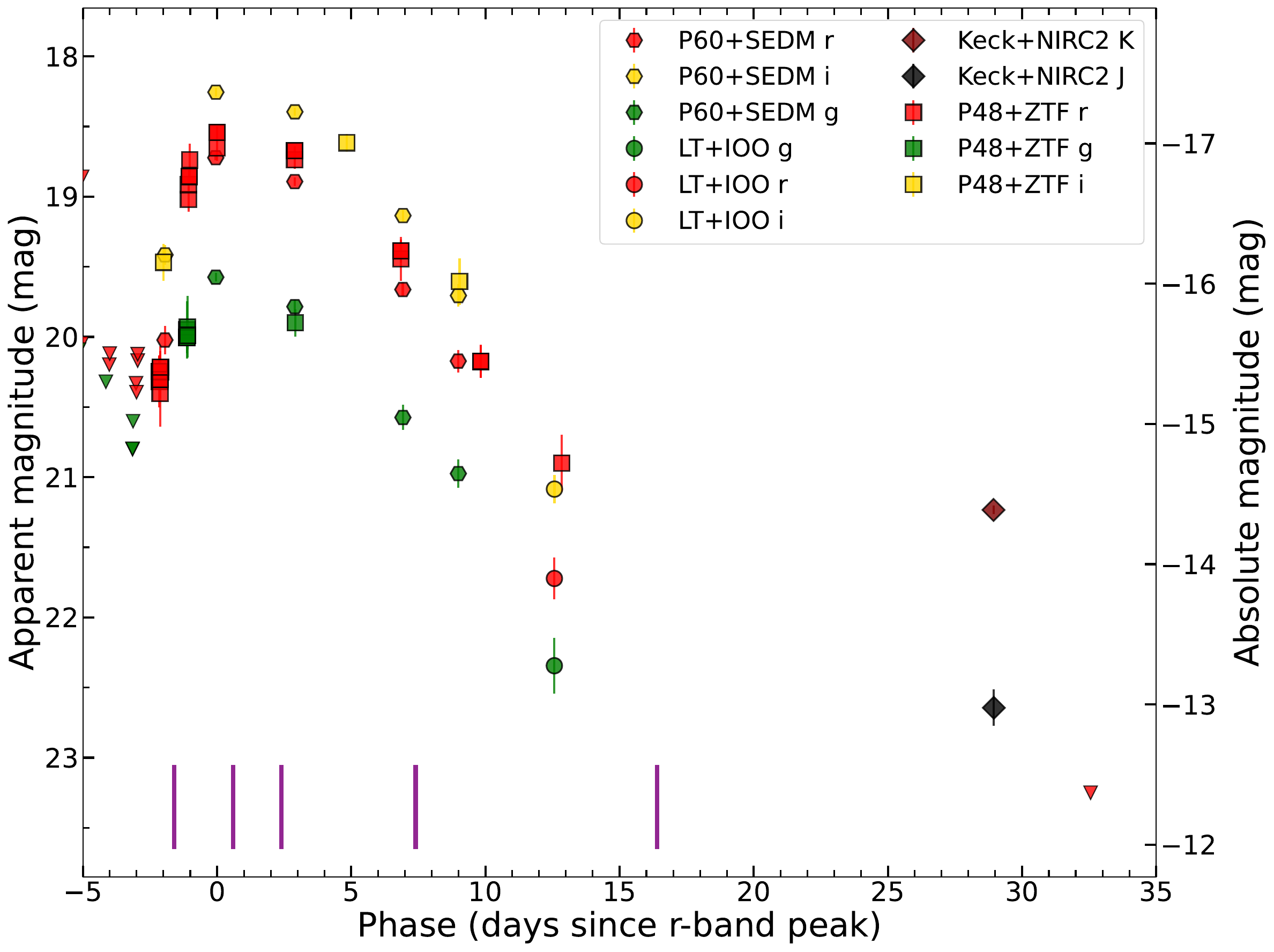}
\caption{Multi-band optical and near-infrared lightcurves of SN~2018erx. The left axis shows the observed apparent magnitudes, while the right axis indicates the corresponding absolute magnitudes. Both axes are corrected for Milky Way extinction, but no correction for host or circumstellar dust extinction has been applied. The phase is given in rest-frame days relative to the epoch of $r$-band maximum. Downward triangles denote nondetections. The epochs of spectroscopic observations are indicated by vertical purple lines.}
    \label{fig:lightcurve}
\end{figure*}

\begin{figure*}
    \centering
    \includegraphics[width=18cm]{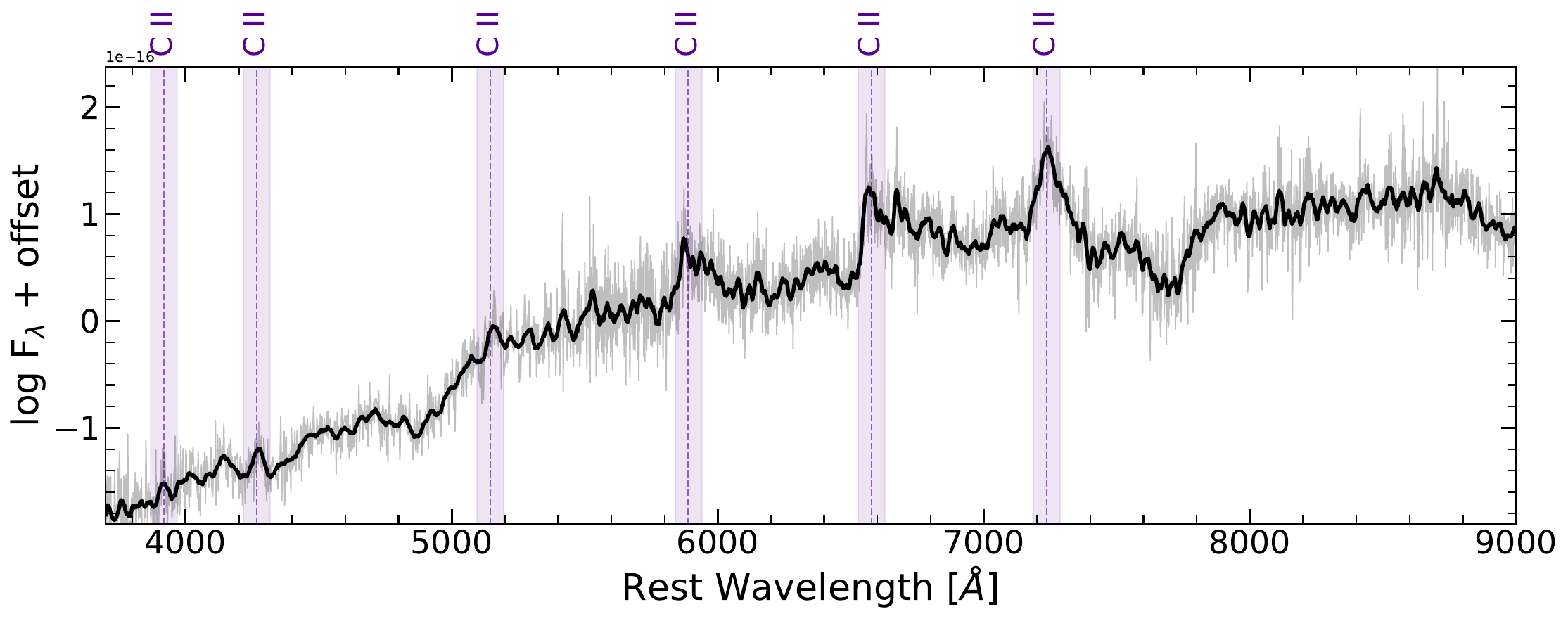}
    \caption{Keck/LRIS spectrum of SN~2018erx obtained at a phase of $+2.4$~days (rest-frame) relative to $r$-band maximum.  The grey line shows the original spectrum, while the black curve indicates a smoothed version. The purple shaded regions mark the wavelengths of prominent \ion{C}{2} $\lambda\lambda$4267, 5145, 5890, 6578, and 7236 emission lines.}
    \label{fig:CIIspectrum}
\end{figure*}

\subsection{Optical and near-infrared photometry}
\label{section:photdata_2018erx}

We obtained multiple epochs of $g$- and $r$-band photometry with the ZTF camera on the Palomar 48-inch Oschin telescope \citep{Dekany20}. The images were processed with the ZTF image analysis pipeline \citep{Masci2019}, and we performed forced photometry at the transient position on the ZTF difference images in the $gri$ bands. We also carried out follow-up optical imaging with the 2.0\,m robotic Liverpool Telescope \citep{Steele2004}, obtaining regular-cadence photometry in the $g$, $r$, and $i$ bands. Further multi-band optical photometry in $g$, $r$, and $i$ was obtained with the Rainbow Camera on the automated Palomar 60-inch telescope \citep[P60;][]{Cenko2006}. The P60 images were processed with the FPipe image-subtraction pipeline \citep{Fremling2016} using Pan-STARRS1 \citep[PS1;][]{Chambers2016} reference images. Near-infrared imaging was obtained in the $J$ and $K$ bands with NIRC2 behind the Keck~II adaptive optics system.\footnote{https://www2.keck.hawaii.edu/inst/nirc2/} The NIRC2 data were reduced using the pipeline described in \citet{De2020c}. The full set of optical and near-infrared lightcurves is shown in Figure~\ref{fig:lightcurve}, and the photometry is listed in Table~\ref{tab:all_phot}. All phases are quoted relative to rest-frame $r$-band maximum (58334.4 MJD) unless otherwise stated.

\subsection{Optical and near-infrared spectroscopy}
\label{section:spectra_2018erx}

We observed the spectral evolution of SN~2018erx, spanning from discovery through the following month. An initial spectrum was taken on the discovery date (2018 August 3) with the Spectral Energy Distribution Machine \citep[SEDM;][]{Blagorodnova2018} on the Palomar 60-inch telescope (with two exposures obtained that night), followed by optical spectroscopy with the Double Spectrograph \citep[DBSP;][]{Oke1982} on the Palomar 200-inch Hale telescope on 2018 August 4. We obtained an additional optical spectrum with the Low-Resolution Imaging Spectrometer \citep[LRIS;][]{Oke1995} on Keck~I on 2018 August 8. Continued monitoring was carried out with DBSP on 2018 August 12 and 2018 August 21. Finally, we obtained a late-time Keck/LRIS spectrum on 2025 August 19 to characterize the host-galaxy contribution at the transient position.

The DBSP spectra were reduced following the procedures described by \citet{Bellm2016} and \citet{Roberson2022}, the SEDM data were reduced using the pipeline described by \citet{Rigault2019}, and the LRIS spectra were reduced using the \lpipe{} pipeline \citep{Perley2019}.

The photometry and spectroscopy were coordinated through the Global Relay of Observatories Watching Transients Happen (\texttt{GROWTH}) Marshal \citep{Kasliwal2019} and the \texttt{Fritz} Marshal \citep{van2019,Coughlin2023}, both of which are web portals designed for vetting and coordinating transient follow-up observations.

Figure~\ref{fig:CIIspectrum} shows an early-time spectrum of SN~2018erx, highlighting the prominent \ion{C}{2} emission features that characterize the event. The spectroscopy log is provided in Table~\ref{tab:spec_log}. All photometry and spectroscopy data will be made publicly available on Zenodo and WISeREP \citep{Yaron:2012aa} after publication.

\section{Extinction correction}
\label{sec:extinction}

We correct all photometry and spectra for Milky Way foreground extinction using the dust maps of \citet{Schlafly11}. Along the line of sight to SN~2018erx we adopt $E(B-V)_{\rm MW}=0.02$~mag. All reddening corrections are applied with the \citet{Cardelli1989} extinction law assuming $R_V=3.1$.

\subsection{Host-galaxy extinction constraints from narrow features}
\label{sec:extinction_host_constraints}

No prominent narrow Na\,\textsc{i} D absorption lines were seen in the spectra of SN~2018erx. To estimate any host-galaxy reddening, we also use the Balmer decrement measured from narrow nebular emission lines in our Keck/LRIS long-slit spectrum taken $\sim$2572~d after $r$-band peak (Fig.~\ref{fig:host}). At this late epoch, any direct contribution from the SN or ongoing SN--CSM interaction should be negligible, so we interpret the narrow Balmer emission as arising from the local host-galaxy environment at the SN position. After shifting the spectrum to the host rest frame ($z=0.0294$), we subtract a locally fitted linear continuum in the $4000$--$5400$~\AA\ region (excluding windows around the Balmer lines) and measure line fluxes for H$\gamma$ and H$\beta$ by integrating the continuum-subtracted spectrum within $\pm 20$~\AA\ of the line centers. We then compare the observed ratio $(\mathrm{H}\gamma/\mathrm{H}\beta)_{\rm obs}$ to the intrinsic Case~B recombination value $(\mathrm{H}\gamma/\mathrm{H}\beta)_0=0.469$, appropriate for photoionized gas at $T\sim 10^4$~K and low electron density \citep[e.g.,][]{Momcheva2013a}. The implied color excess in AB mag is
\begin{equation}
E(B-V)_{\rm gas} = \frac{2.5}{k(\mathrm{H}\gamma)-k(\mathrm{H}\beta)}
\log_{10}\left[\frac{(\mathrm{H}\gamma/\mathrm{H}\beta)_0}{(\mathrm{H}\gamma/\mathrm{H}\beta)_{\rm obs}}\right],
\end{equation}
where $k(\lambda)=A_\lambda/E(B-V)$ is evaluated from the \citet{Cardelli1989} law. For $R_V=3.1$ we adopt $k(\mathrm{H}\beta)=4.60$ and $k(\mathrm{H}\gamma)=5.12$, and convert to visual attenuation via $A_V^{\rm host}=R_V\,E(B-V)_{\rm gas}$. We measure $(\mathrm{H}\gamma/\mathrm{H}\beta_{\rm obs}=0.459$, consistent with the intrinsic Case~B value within uncertainties, implying no significant host-galaxy reddening. 

We therefore find no compelling evidence for significant large-scale host-galaxy extinction from either narrow \ion{Na}{1}~D absorption or the Balmer decrement. However, Balmer-decrement estimates can be affected by underlying stellar Balmer absorption, slit losses, residual calibration uncertainties, and the narrow nebular lines need not trace the same dust column that attenuates the transient continuum if the emitting gas is not strictly co-spatial with the SN or if additional dust is local to the progenitor system. Likewise, extinction estimates based on the equivalent width of narrow \ion{Na}{1}~D are uncertain, particularly in low-resolution spectra \citep{Poznanski2011} and in the presence of circumstellar material \citep{Phillips2013}.

\subsection{Effective reddening implied by the red continuum and early ionized-carbon features}
\label{sec:extinction_carbon_summary}

Blackbody fits to the early-time optical SED of SN~2018erx, corrected only for Milky Way extinction, yield characteristic temperatures of only $\sim4000$~K if no additional extinction is applied (see Section \ref{sec:bbfit}). Such low temperatures are difficult to reconcile with the early spectroscopic appearance of the SN, which shows prominent \ion{C}{2} features and a Type~Icn-like interaction spectrum. We therefore interpret the observed red optical colors as reflecting substantial {effective} attenuation of an intrinsically hotter continuum, rather than the true color temperature of the emitting region.

This interpretation is also supported by comparison with other Type~Icn SNe. Published events with analogous early-time ionized-carbon features show blue continua and characteristic blackbody temperatures of order $10^{4}$~K or higher at comparable phases \citep{Pellegrino2022,perley2022,Nagao2023,Davis2023}, while radiative-transfer models of hot carbon-rich stripped-envelope explosions produce optical \ion{C}{2} features near maximum light at photospheric temperatures of $\sim12,000$~K \citep{Dessart2019}. Motivated by the temperatures inferred for Type~Icn SNe and by spectral models that produce \ion{C}{2} features at $T\sim12,000$~K, we adopt a fiducial intrinsic continuum temperature of $T_{\rm C}=12,000$~K and a conservative range of $10{,}000$--$15,000$~K.

Close to the epoch of the MJD~58333.8 LRIS spectrum, the absolute magnitudes corrected for MW extinction are $M_g=-15.8$, $M_r=-16.7$, and $M_i=-17.2$, which give observed colors of $(M_g-M_r)_{\rm obs}=0.9$~mag and $(M_r-M_i)_{\rm obs}=0.5$~mag. We approximate the intrinsic optical continuum by a blackbody with color temperature $T_{\rm C}$ and evaluate flux ratios at the effective wavelengths of the $gri$ filters, $\lambda_g = 0.477\ \mu{\rm m}$, $\lambda_r = 0.623\ \mu{\rm m}$, and $\lambda_i = 0.763\ \mu{\rm m}$, using the Planck function:
\begin{equation}
B_\nu(\nu,T) = \frac{2h\nu^3}{c^2}\,\frac{1}{\exp(h\nu/kT)-1}.
\label{eq:planck}
\end{equation}

In the AB system, the monochromatic blackbody color between two bands $X$ and $Y$ is
\begin{equation}
(X-Y)_{\rm BB}(T_{\rm C}) \simeq -2.5\log_{10}\!\left[\frac{B_\nu(\nu_X,T_{\rm C})}{B_\nu(\nu_Y,T_{\rm C})}\right],
\label{eq:bb_color}
\end{equation}
where $\nu_X = c/\lambda_X$.

Across the adopted temperature range, the corresponding blackbody colors are $(M_g-M_r)_{\rm BB}\approx -0.04$ to $-0.25$~mag and $(M_r-M_i)_{\rm BB}\approx -0.14$ to $-0.25$~mag. Adopting a Cardelli extinction law with $R_V=3.1$, the relevant color coefficients are $R_g-R_r = 1.005$ and $R_r-R_i = 0.673$. Matching the de-reddened colors to the blackbody colors gives
\begin{equation}
E(B-V)_{gr} = \frac{(M_g-M_r)_{\rm obs} - (M_g-M_r)_{\rm BB}}{1.005},
\end{equation}
\begin{equation}
E(B-V)_{ri} = \frac{(M_r-M_i)_{\rm obs} - (M_r-M_i)_{\rm BB}}{0.673}.
\end{equation}

Across this temperature range, both color combinations yield similar values,
\begin{equation}
E(B-V) \simeq 1.0\text{--}1.2~{\rm mag}.
\end{equation}

Using $T_{\rm C}=12{,}000$~K as a fiducial intrinsic continuum temperature, we obtain an effective reddening estimate of $E(B-V)\simeq 1.05$~mag, corresponding to $A_V \simeq 3.3$~mag for $R_V=3.1$. This estimate is robust to within $\pm0.1$~mag in $E(B-V)$, corresponding to $\pm0.3$~mag in $A_V$, across the adopted temperature range.

\subsection{Consistency with CSM-interaction lightcurve modeling}
\label{sec:extinction_redback_consistency}

As an independent check, we allow the effective extinction to vary freely in our multi-band CSM-interaction modeling with \texttt{redback} \citep{sarin_redback,Chatzopoulos2012,Chatzopoulos2013} (see Sect.~\ref{sec:lcmodelling}). In this framework, the model SED is approximated as a blackbody (with a temperature floor) and an attenuation parameter $A_V^{\rm eff}$ is inferred simultaneously with the interaction parameters. The posterior favors substantial attenuation,
\begin{equation}
A_V^{\rm eff} \simeq 3.15^{+0.05}_{-0.05}~{\rm mag},
\end{equation}
consistent with the \ion{C}{2}-motivated color-temperature estimate. Given the strong and internally consistent evidence for significant effective attenuation from the early-time colors and the CSM-interaction fit, we adopt $A_V^{\rm eff}\simeq 3.3$~mag, together with $E(B-V)_{\rm MW}=0.02$~mag, for extinction corrections throughout this work. This choice also brings the color evolution of SN~2018erx into agreement with that of other interacting SESNe (Figure~\ref{fig:color}).

\subsection{Consistency with the inferred dust properties}
\label{sec:extinction_dust_consistency}

In the previous subsections, we inferred a large effective attenuation,
$A_V^{\rm eff}\simeq 3.2$--$3.3$~mag, from the lightcurve and the presence of \ion{C}{2} features, despite the absence of strong host-ISM reddening diagnostics (narrow \ion{Na}{1}~D absorption and the Balmer decrement). This motivates a local circumstellar dust origin for much of the extinction. Here we assess whether an attenuation of this magnitude is quantitatively consistent with the dust component implied by the observed NIR excess.

For a foreground dust screen, extinction and optical depth are related by
\begin{equation}
A_V = 2.5\log_{10}(e)\,\tau_V \simeq 1.086\,\tau_V,
\end{equation}
 where $\tau_V$ is the $V$-band extinction optical depth (absorption + scattering). Writing $\tau_V=\kappa_V \Sigma_{\rm d}$ in terms of the dust surface density $\Sigma_{\rm d}$ and the $V$-band mass extinction coefficient $\kappa_V$ (e.g., \citealt{Draine2011}), and approximating the dust as residing in a geometrically thin shell at radius $R_{\rm d}$ that subtends a solid-angle covering fraction $f_\Omega$ of $4\pi$, the characteristic dust column through the dusty regions is
\begin{equation}
\Sigma_{\rm d}\simeq \frac{M_{\rm d}}{4\pi f_\Omega R_{\rm d}^2}.
\end{equation}
Combining these relations gives
\begin{equation}
A_V \simeq 1.086\,\kappa_V\,\frac{M_{\rm d}}{4\pi f_\Omega R_{\rm d}^2}.
\label{eq:av_shell}
\end{equation}

Solving for the required dust mass yields
\begin{equation}
\label{eq:md_req_smallR}
\begin{aligned}
M_{\rm d} \simeq\;&
8.6\times10^{-6}\,
\left(\frac{A_V}{3.3~\rm{mag}}\right)
\left(\frac{R_{\rm d}}{3\times10^{15}\ {\rm cm}}\right)^{2} \\
&\times
\left(\frac{\kappa_V}{2\times10^{4}\ {\rm cm^2\,g^{-1}}}\right)^{-1}
\left(\frac{f_\Omega}{1}\right)
\,M_\odot ,
\end{aligned}
\end{equation}
where the normalization $\kappa_V\simeq 2\times10^{4}\ {\rm cm^2\,g^{-1}}$ is representative of carbonaceous (see Section~\ref{sec:dust}) ISM-like grain populations at optical wavelengths (e.g., \citealt{Draine2003,Zubko2004}).

Equation~\eqref{eq:md_req_smallR} shows that if the extinction-producing dust lies at a radius of order a few $\times10^{15}$~cm (comparable to the effective emitting scale inferred for the hot NIR component in Section~\ref{sec:dust}), only $\sim10^{-6}$--$10^{-5}\,M_\odot$ of dust is required to produce $A_V\sim3$~mag for plausible optical opacities. This is comparable to the order-of-magnitude dust mass inferred from the $JK$ emission under optically thin assumptions (Section~\ref{sec:dust}) and to the lower end of dust masses inferred in other interacting stripped-envelope SNe \citep[e.g.,][]{Anupama2009,Gan2021,Yamanaka2025}. An extinction of $A_V\sim3$~mag is therefore not extreme from a dust-mass perspective. More generally, the required dust mass scales as $M_{\rm d}\propto R_{\rm d}^2\,f_\Omega/\kappa_V$, so a clumpy or aspherical distribution ($f_\Omega<1$) can yield substantial line-of-sight attenuation without requiring a globally massive dust shell. Taken together, the NIR-inferred dust properties and the color-based extinction estimate are mutually consistent with a circumstellar dust origin for the attenuation.

\section{Observed properties}
\label{sec:observables}

\subsection{Explosion epoch estimation}\label{sec:explosion}


For SN~2018erx, we constrain the explosion epoch using the best-fitting CSM interaction model (see Section \ref{sec:analysis}), which yields $t_{\rm exp} = 58328.19^{+0.04}_{-0.04}\ {\rm MJD}$. The ZTF lightcurve includes a last $r$-band non-detection at MJD~58331.17 with a limiting magnitude of $r>20.4$~mag, followed by the first $r$-band detection at MJD~58331.18 with $r=20.3$~mag in the ZTF forced-photometry light curve. 

\subsection{Lightcurve properties of SN~2018erx}

\begin{figure*}
    \centering

    \includegraphics[width=9cm]{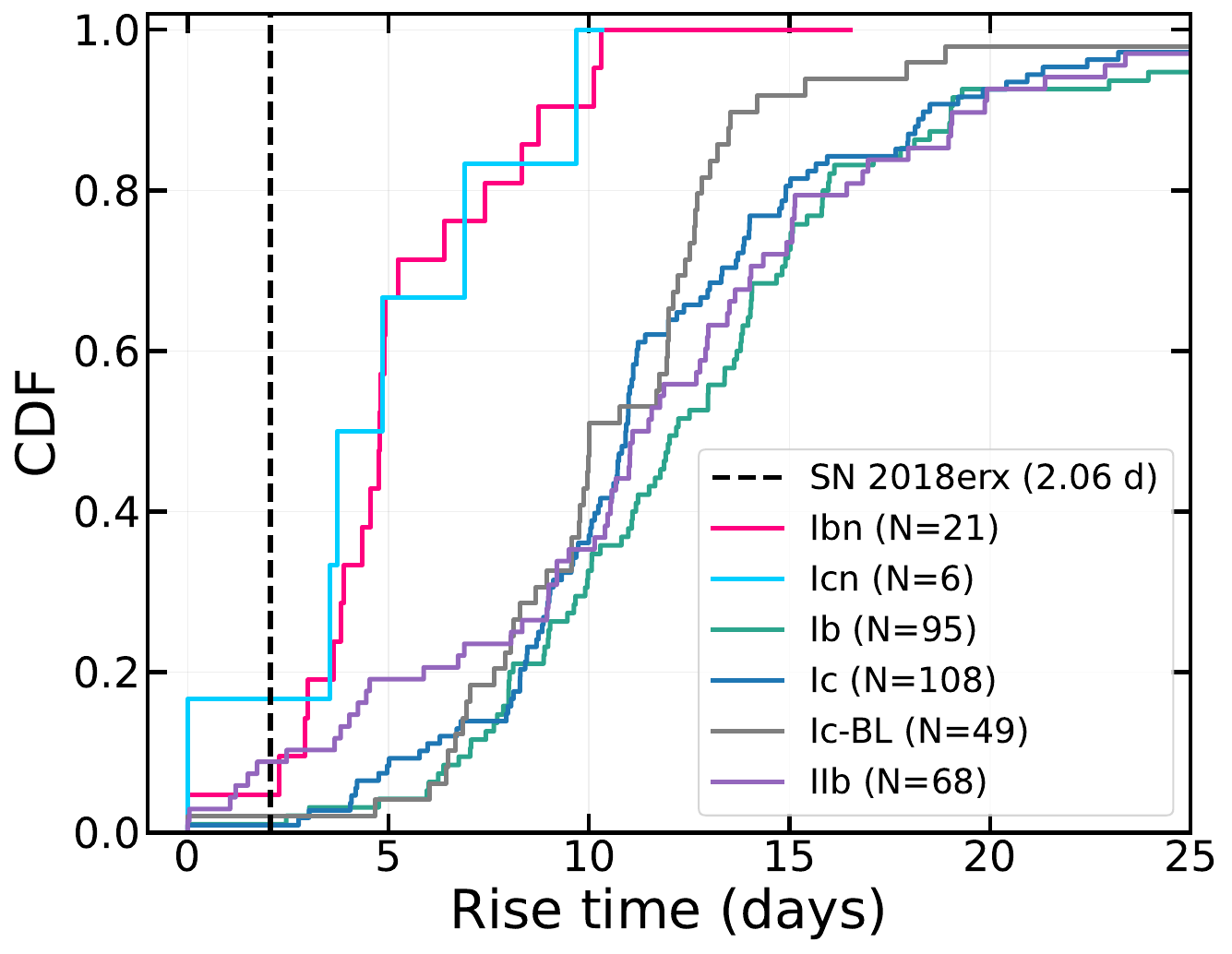}\includegraphics[width=9cm]{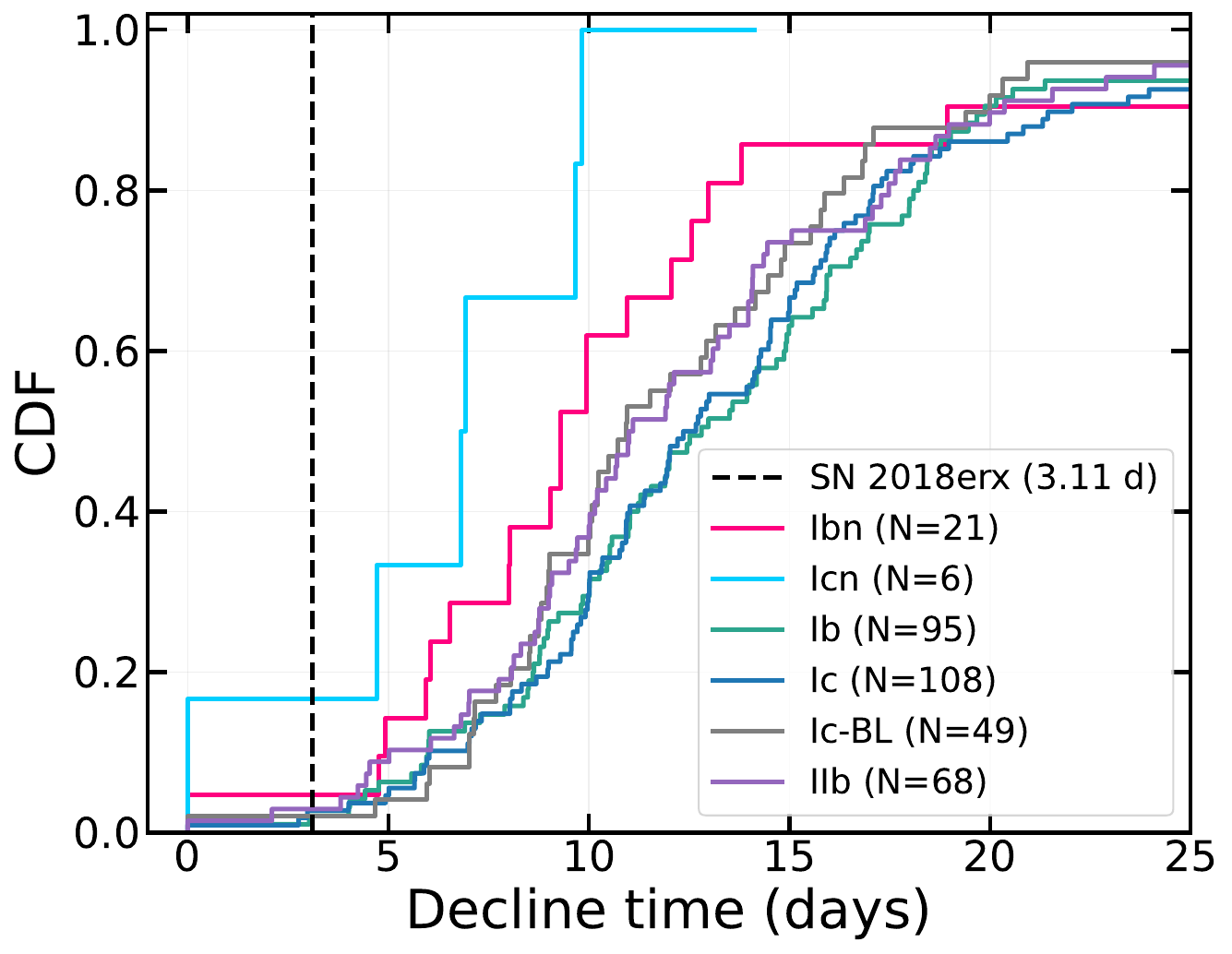}
    \caption{Cumulative distribution functions (CDFs) of rise time (left) and decline time (right) for stripped-envelope SN subclasses from the ZTF Bright Transient Survey (BTS) sample, while the SNe~Icn sample combines the BTS SNe~Icn with SN~2019jc, SN~2021ckj, SN~2023qre, and SN~2023rau. The vertical dashed line indicates the measured values for SN~2018erx, showing that it lies among the most rapidly evolving events relative to the broader populations of SNe~Ib, Ic, Ic-BL, and IIb. 
}
    \label{fig:timescale}
\end{figure*}

The multi-band light curve of SN~2018erx, shown in Figure~\ref{fig:lightcurve}, spans $\sim$$-3$ to $+30$~d relative to $r$-band maximum. To estimate the peak date and magnitude, we interpolate the photometry with a Gaussian process implemented in \texttt{GEORGE} \citep{Ambikasaran2015} using a Mat\'ern-3/2 kernel. The GP fit yields an $r$-band maximum at MJD~58334.4 with $r=18.32\pm0.17$~mag. 
We measure the light-curve width in flux space relative to half-maximum: the rest-frame rise time from 50\% of peak flux to maximum is 2.1~d, and the corresponding fade time from maximum back to 50\% of peak flux is 3.1~d. The rapid post-peak evolution is also reflected in $\Delta m_{15,r}=2.9$~mag. These values place SN~2018erx among the fastest-evolving events in the SESN and Type Ibn/Icn populations \citep{Hosseinzadeh2017,Farias2025,Pellegrino2022b}, 
and comparable to other rapidly evolving stripped-envelope transients \citep{Drout2014}. This is illustrated in Figure~\ref{fig:timescale}, where SN~2018erx occupies a short-timescale region of parameter space relative to the broader stripped-envelope SN population. The estimated $r$-band decline rates over the intervals 0--10 and 10--20~d after maximum are $0.16\pm0.01$ and $0.09\pm0.01$~mag~d$^{-1}$, respectively, faster than the mean weighted decline rate of the SN Ibn sample in \citet{Farias2025}. The cumulative distributions shown in Figure~\ref{fig:timescale} also indicate that SN~2018erx is among the most rapidly evolving events relative to the Bright Transient Survey (BTS) populations of SNe~Ib, Ic, Ic-BL, and IIb. In addition to the unusually rapid optical evolution, SN~2018erx also exhibits a pronounced late-time near-infrared excess. At $\sim$$+29$~d, the transient is detected in both the $J$ and $K$ bands with very red $(J-K)$ color, indicating the presence of a hot dust component. We analyze the inferred dust properties in Section~\ref{sec:dust}.

\begin{figure*}
    \centering
    \includegraphics[width=\textwidth]{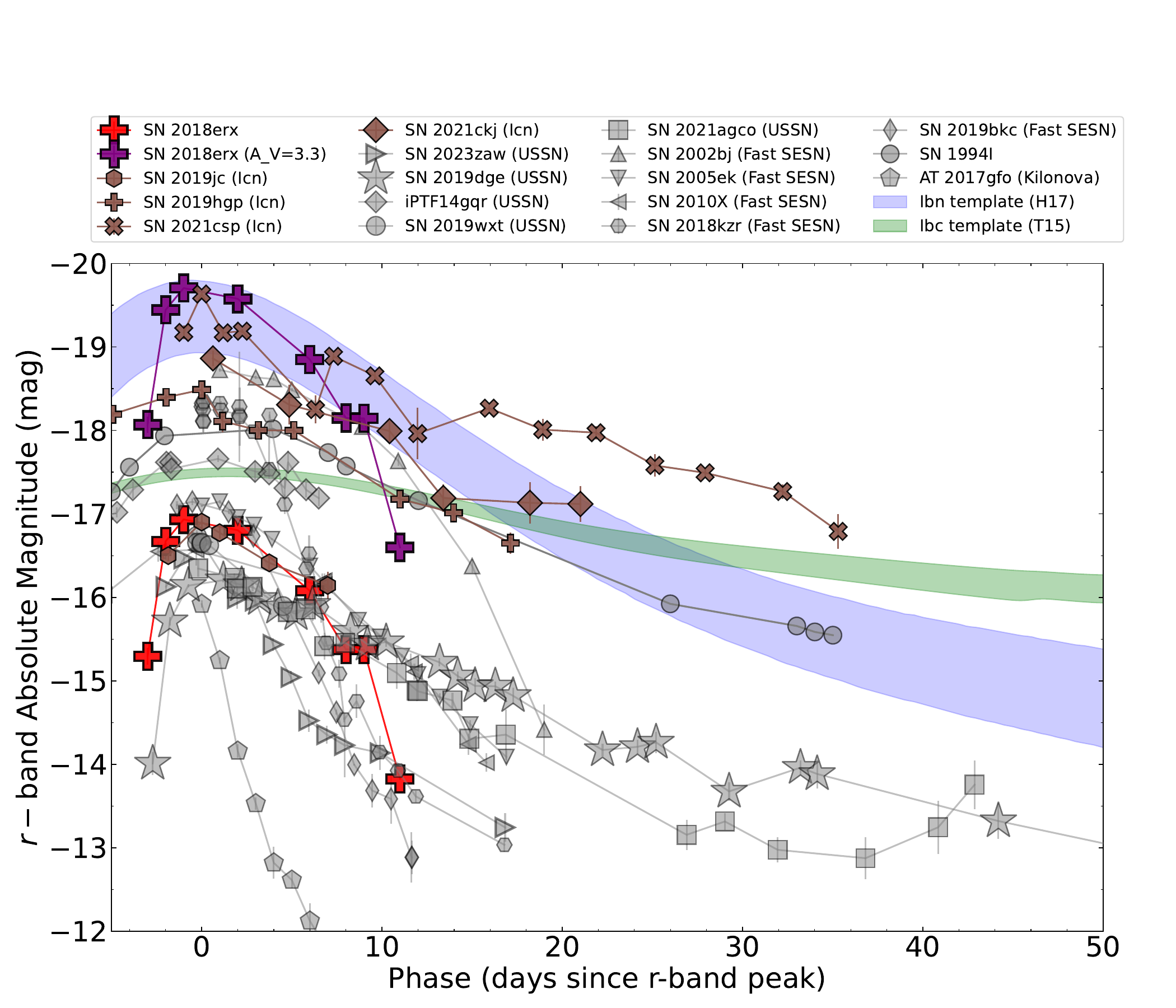}
    \caption{$r$-band absolute-magnitude lightcurve of SN~2018erx compared with rapidly evolving transients and SESNe from the literature. Red symbols show SN~2018erx assuming no host extinction ($A_V=0$), while purple symbols show the extinction-corrected lightcurve adopting $A_V=3.3$~mag. For comparison, lightcurves of several fast transients and interacting SESNe (colored markers) and a broader sample of stripped-envelope SNe (gray markers) are shown. The shaded regions indicate template lightcurves for SNe~Ibn \citep[G17;][]{Hosseinzadeh2017} and SNe~Ibc \citep[T15;][]{Taddia2015}. 
    SN~2018erx evolves on a significantly faster timescale than typical SESNe. 
}
    \label{fig:phot_all}
\end{figure*}

\subsection{Comparison of the lightcurve with literature transients}

Using the distance adopted in Section~\ref{sec:discovery_2018erx}, we construct the absolute-magnitude $r$-band light curve of SN~2018erx and compare it with rapidly evolving transients and stripped-envelope SNe from the literature (Figure~\ref{fig:phot_all}).  We further summarize its location in peak-luminosity versus timescale space in Figure~\ref{fig:timescale}. 
To place SN~2018erx in context, our comparison sample includes rapidly evolving hydrogen-poor transients such as SNe~2002bj \citep{Poznanski2010}, 2005ek \citep{Drout2013}, 2010X \citep{Kasliwal2010}, 2018kzr \citep{McBrien2019}, 2019dge \citep{Yao2020}, iPTF14gqr \citep{De2018c}, 2019bkc \citep{Prentice2020,Chen2020}, 2019wxt \citep{Agudo2023,Shivkumar2023}, and 2021agco \citep{Yan2023}, together with the well-studied Type~Ic SN~1994I \citep{Richmond1996} and representative template light curves of SNe~Ibn and SNe~Ibc \citep{Hosseinzadeh2017,Taddia2015}. Several of these objects, in particular SNe~2002bj, 2005ek, 2010X, 2019wxt, and 2021agco, occupy a similar fast-timescale regime. Motivated by our interaction modeling, which favors a low ejecta mass, we also compare SN~2018erx with transients proposed to arise from ultra-stripped explosions, including SNe~2019dge \citep{Yao2020}, 2023zaw \citep{Das2024}, 2021agco \citep{Yan2023}, and iPTF14gqr \citep{De2018c}. For completeness, we further include AT2017gfo, the electromagnetic counterpart of GW170817 \citep{Abbott2017,Villar2017}, which is not a core-collapse event but provides a useful benchmark for extremely rapid photometric evolution.

We estimate the $r$-band peak using a Gaussian-process interpolation with a Mat\'ern-3/2 kernel, obtaining an observed peak absolute magnitude of $M_r=-17.29\pm0.17$~mag. Prior to any host-extinction correction, SN~2018erx would therefore be among the faintest members of the interacting SESN population and is roughly two magnitudes fainter than the mean SN~Ibn/Icn templates, despite having similarly rapid rise and decline times. Adopting the effective host attenuation inferred in Section~\ref{sec:extinction}, $A_V=3.3$~mag, and a \citet{Cardelli1989} law with $R_V=3.1$, the corresponding extinction in the $r$ band is $A_r\simeq2.87$~mag. This shifts the intrinsic peak to $M_r\approx-20.16$~mag. Thus, while the uncorrected light curve would place SN~2018erx among the faintest interacting SESNe, the extinction-corrected luminosity moves it into the luminous end of the Ibn/Icn-like population, while preserving its exceptionally short timescale. 

\begin{figure*}
    \centering
    \includegraphics[width=14.5cm]{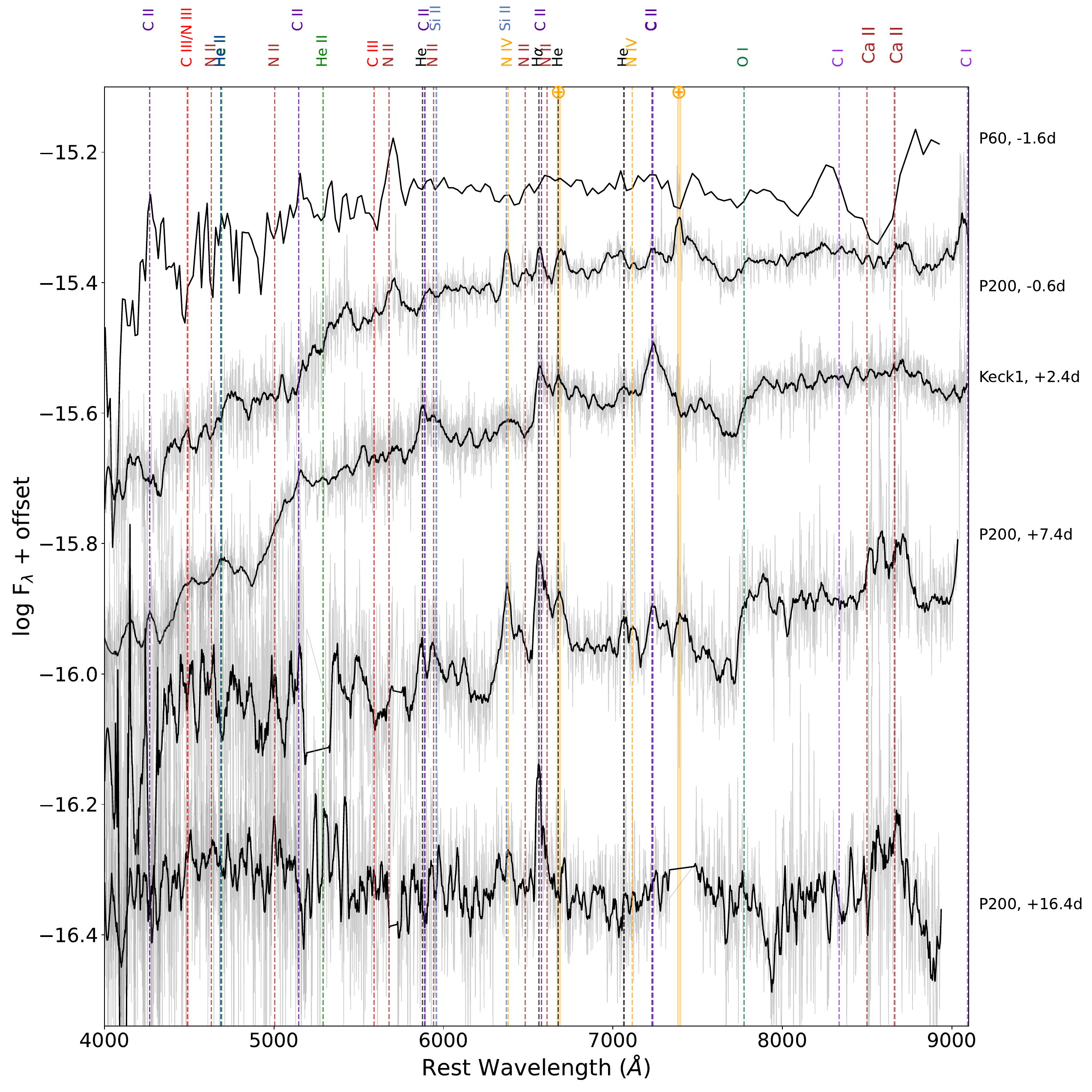}
    \caption{Spectral evolution of SN~2018erx from $-1.6$ to $+16.4$ days relative to $r$-band maximum. 
The spectra are shown in the host-galaxy rest frame and have been vertically offset for clarity. 
The gray lines show the original spectra, while the black curves indicate smoothed versions. 
Indentifications of prominent lines are marked with dashed vertical lines, including \ion{C}{2}, \ion{He}{1}, \ion{N}{2}, \ion{N}{3}, \ion{O}{1}, and Ca\,\textsc{ii}. }
    \label{fig:spec_all}
\end{figure*}

\subsection{Optical and near-infrared spectral properties of SN~2018erx}
\label{sec:spectralevolution}

\label{section:spectra_prop_2018erx}


Figure~\ref{fig:spec_all} shows the spectroscopic evolution of SN~2018erx from $-1.6$ d to $+16.4$ d relative to $r$-band maximum. The spectra shown are corrected for the host redshift but are not corrected for extinction from the Milky Way or the host galaxy along the line of sight. The earliest spectra exhibit a remarkably red continuum. Such a red continuum could arise either from substantial dust extinction along the line of sight or from an intrinsically red spectral energy distribution \citep{Sasdelli2016}.

No unambiguous signatures of flash ionization are identified in the early spectra. Rather than exhibiting the rapidly fading, transient emission features typically associated with flash-ionized infant supernovae, the spectra obtained between $-1.6$ d and $+2.4$ d show comparatively long-lived absorption and emission features of C\,\textsc{ii}, N\,\textsc{ii}, Si\,\textsc{ii}, and O\,\textsc{i}, which strengthen with time. Weak He\,\textsc{i} $\lambda5876$ and $\lambda6678$ emission is also detected over this interval. The prevalence of lower-ionization species may point either to a cooler ejecta environment, consistent with the blackbody temperature evolution shown in Fig.~\ref{fig:blackbody}, or to a softer ionizing continuum.

At later epochs, from $+2.4$ d to $+16.4$ d, the C\,\textsc{ii} features strengthen significantly, particularly the transitions at $\lambda5890$, $\lambda6578$, and $\lambda7236$. Lines of N\,\textsc{ii} and O\,\textsc{i} remain visible throughout this period, while the Ca\,\textsc{ii} near-infrared triplet becomes progressively stronger with time. The overall strengthening of C\,\textsc{ii} features, together with the rapid photometric evolution (Section~\ref{section:photdata_2018erx}), supports a classification of SN~2018erx as a Type~Icn SN. An important characteristic of SN~2018erx is the detection of He\,\textsc{i} lines at multiple stages of its evolution. This distinguishes it from previously studied SNe~Icn, including SN~2019hgp and SN~2021csp, which did not show clear He\,\textsc{i} features. The presence of helium in SN~2018erx thus points to additional diversity within the SN~Icn population.

\begin{figure}
    \centering
    \includegraphics[width=9.0cm]{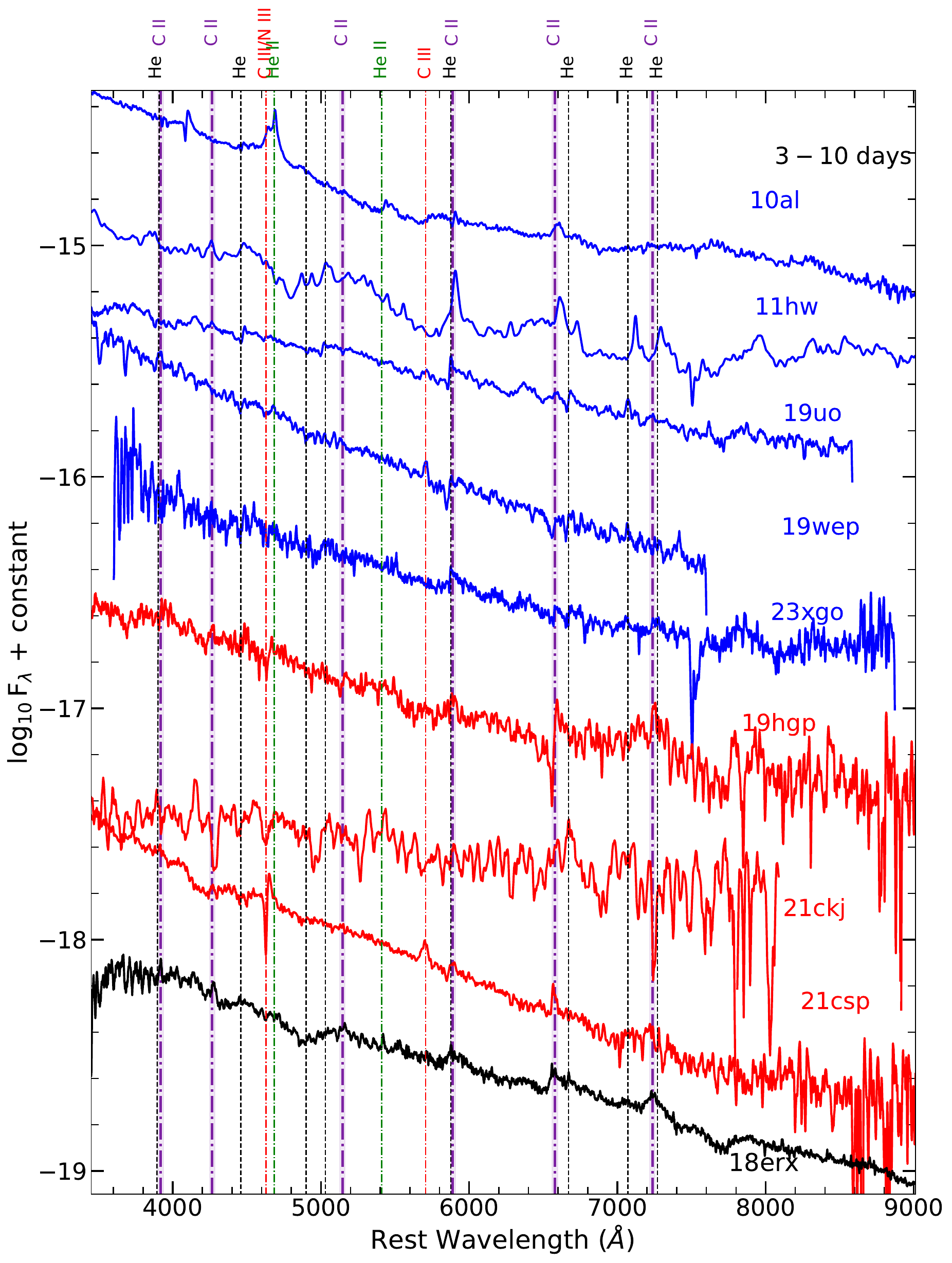}
    \caption{Comparison of the +2.4~day Keck/LRIS spectrum of SN~2018erx (black) with representative spectra of Type~Ibn (blue) and Type~Icn (red) supernovae at phases of $\sim3$--10 days relative to peak. 
The spectra are shown in the rest frame and vertically offset for clarity. 
Dashed vertical lines mark the wavelengths of prominent spectral features, including \ion{C}{2}, \ion{He}{1}, \ion{He}{2}, \ion{N}{2}, \ion{N}{3}, and Ca\,\textsc{ii}. }
    \label{fig:spectracompare}
\end{figure}

\begin{figure}
    \centering
    \includegraphics[width=9.0cm]{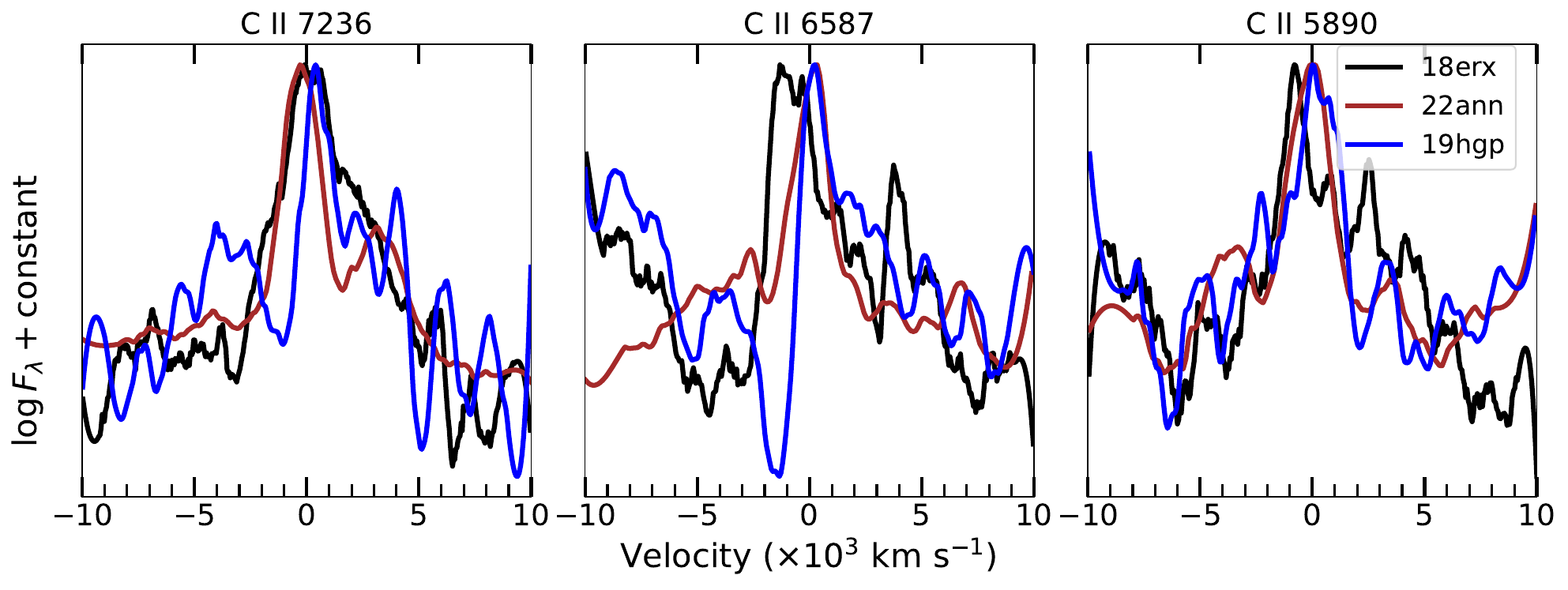}
    \caption{
Velocity profiles of prominent \ion{C}{2} emission lines in SN~2018erx compared with representative Type~Icn supernovae. 
The panels show the regions around \ion{C}{2} $\lambda$7236, $\lambda$6587, and $\lambda$5890. 
Spectra are plotted in velocity space relative to the rest wavelength of each transition. The spectrum of SN~2018erx (black) is compared with those of the Type~Icn events SN~2022ann (red) and SN~2019hgp (blue). The similar line widths and profiles highlight the strong \ion{C}{2} emission characteristic of carbon-rich interacting SESNe.
}
    \label{fig:CII}
\end{figure}

\subsection{Comparison of spectra with literature transients}

To further assess the spectral classification of SN~2018erx, we compare its earliest high-quality optical spectrum, the $+2.4$~d Keck/LRIS spectrum, with representative Type~Ibn and Type~Icn events (Figure~\ref{fig:spectracompare}). Among the SNe~Ibn, we consider SNe~2010al \citep{PastorelloSN2010al}, 2011hw \citep{PastorelloSN2010al}, 2019uo \citep{Gangopadhyay2020}, and 2019wep \citep{Gangopadhyay2022}. For the Type~Icn comparison sample, we use SNe~2019hgp \citep{2019hgp}, 2021ckj \citep{Nagao2023}, and 2021csp \citep{PerleyIcn}. All comparison spectra are 
corrected for redshift, and where available they are also corrected for the adopted host extinction. 

The overall comparison favors a Type~Icn classification for SN~2018erx. Its spectrum is dominated by prominent \ion{C}{2} emission features, with clear detections at $\lambda\lambda$4629, 5145, 5706, 5890, 6578, and 7236, and shows the closest resemblance to the Type~Icn SN~2021csp. In particular, the strong carbon features and the absence of the narrow, helium-dominated emission spectrum typical of classical SNe~Ibn argue that the CSM interacting with SN~2018erx is H/He-poor and carbon rich. Weak \ion{He}{1} $\lambda5876$ and $\lambda6678$ are present, but they do not dominate the spectrum. We therefore classify SN~2018erx as a Type~Icn supernova, while noting that it may retain some residual helium seen in other SNe~Icn \citep{PerleyIcn}.

This classification is further supported by the velocity-space comparison of the strongest \ion{C}{2} lines (Figure~\ref{fig:CII}). We compare the profiles of \ion{C}{2} $\lambda7236$, $\lambda6587$, and $\lambda5890$ in SN~2018erx with those of the Type~Icn events SN~2022ann and SN~2019hgp. The line profiles of SN~2018erx are broadly similar in both width and shape to those of the comparison Type Icn events, reinforcing the interpretation that its intermediate-width carbon emission arises from interaction with a carbon-rich CSM. Compared to SN~2021csp, whose \ion{C}{2} lines display narrow P-Cygni structure, the \ion{C}{2} features in SN~2018erx are smoother and somewhat broader, with characteristic FWHM of order $\sim$3500--4000~km~s$^{-1}$. \textcolor{black}{We do not detect the very narrow component of C\,\textsc{ii}, typically observed in interacting SNe and tracing slow unshocked CSM ($v \lesssim 100$--$200$~km~s$^{-1}$). This may be attributed to either insufficient spectral resolution to resolve such a feature, rapid photoionization of the pre-shock CSM by the intense SN radiation field, or a geometrically confined and tenuous circumstellar environment in which narrow-line emission falls below the detection threshold and sometimes can be caused by radiative acceleration suppressing narrow lines.}

\section{Analysis and Results}\label{sec:analysis}

\begin{figure}
    \centering
    \includegraphics[width=9cm]{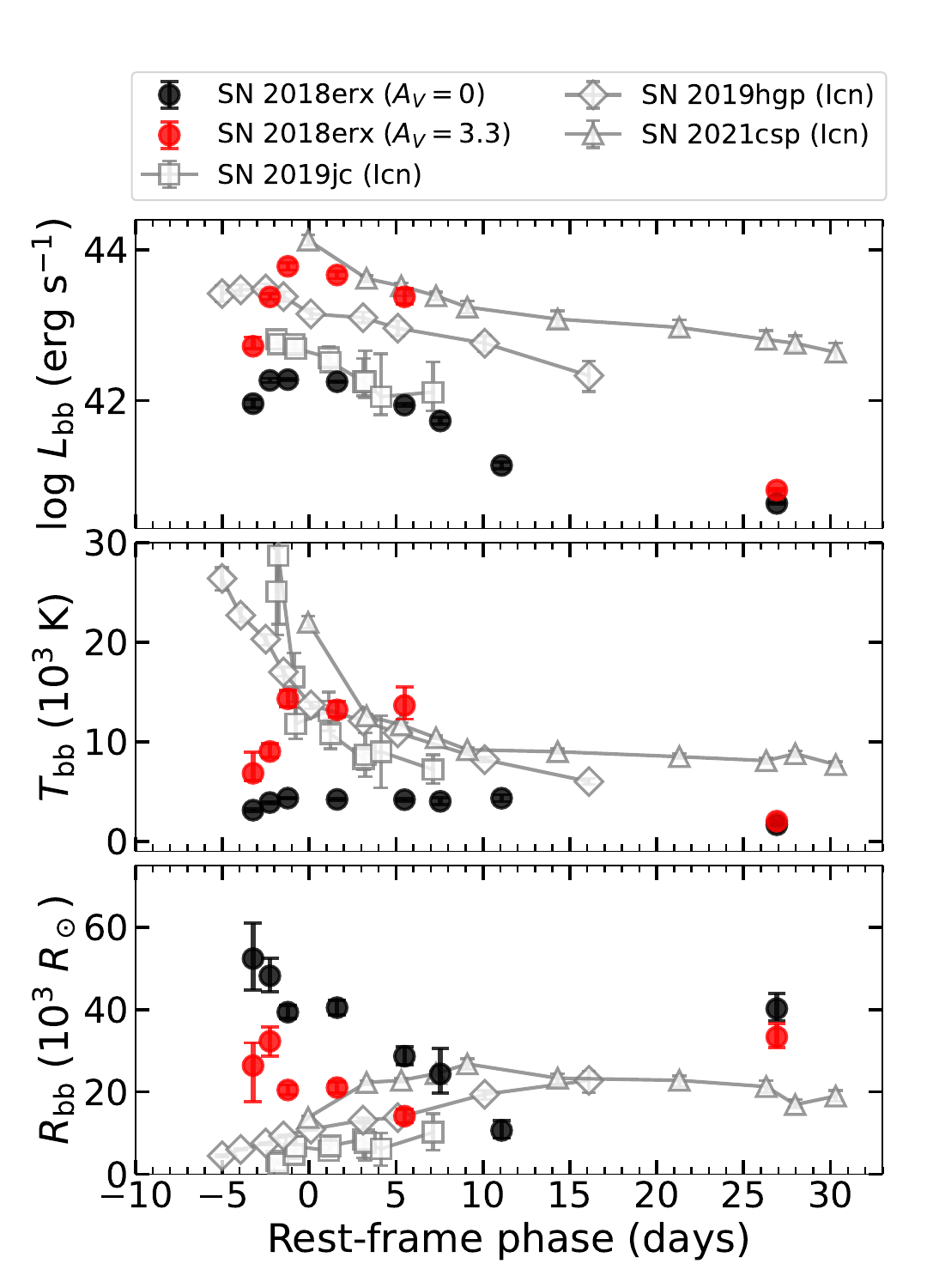}
    \caption{Evolution of the blackbody parameters derived from multi-band photometry of SN~2018erx. 
Top: bolometric luminosity ($L_{\rm bb}$); middle: blackbody temperature ($T_{\rm bb}$); bottom: blackbody radius ($R_{\rm bb}$) as a function of rest-frame phase relative to $r$-band maximum. 
Black symbols show results obtained assuming no host extinction ($A_V=0$), while brown symbols correspond to extinction-corrected values adopting $A_V=3.3$~mag. 
Correcting for the inferred host attenuation significantly increases the inferred temperature and luminosity while reducing the emitting radius near peak.}
    \label{fig:blackbody}
\end{figure}

\subsection{Blackbody SED fitting}
\label{sec:bbfit}
We construct spectral energy distributions (SEDs) by binning the multi-band photometry into 1-day (rest-frame) intervals. For each epoch with detections in at least three filters, we fit a Planck blackbody function to the SED and infer the effective temperature ($T_{\rm bb}$) and photospheric radius ($R_{\rm bb}$). We sample the posterior distribution using the Python package \texttt{emcee} \citep{Foreman-Mackey13}. The corresponding bolometric luminosity is then computed as
\begin{equation}
L_{\rm bb}=4\pi R_{\rm bb}^{2}\sigma T_{\rm bb}^{4}.
\end{equation}
where $\sigma$ denotes the Stefan--Boltzmann constant.
We report the median posterior values in Table~\ref{tab:bbtable}, with uncertainties corresponding to the 16th and 84th percentiles.

Figure~\ref{fig:blackbody} shows the temporal evolution of the inferred blackbody parameters for SN~2018erx, while the corresponding numerical values are listed in Table~\ref{tab:bbtable}. Black symbols show results obtained assuming no host extinction ($A_V=0$), while brown symbols correspond to extinction-corrected values adopting $A_V=3.3$~mag. Correcting for the inferred host attenuation significantly increases the inferred temperature and luminosity while reducing the emitting radius near peak.

Without any host-extinction correction, SN~2018erx reaches a peak bolometric luminosity of $\log(L_{\rm bb}/{\rm erg\,s^{-1}})\approx42.3$ near maximum light. The luminosity then declines rapidly to $\log(L_{\rm bb}/{\rm erg\,s^{-1}})\approx41.1$ by $\sim$10~d after peak and to $\approx40.6$ by the latest epoch at $\sim$27~d. In this case the inferred temperature remains relatively cool, with $T_{\rm bb}\approx(4.0$--$4.5)\times10^{3}$~K during the first $\sim$10~d after peak, before dropping to $\sim1.6\times10^{3}$~K at the final epoch. The corresponding radius is large around maximum, $R_{\rm bb}\sim(4$--$6)\times10^{4}\,\Rsun$, decreases to $\sim10^{4}\,\Rsun$ by $\sim$+11~d, and remains of order a few $\times10^{4}\,\Rsun$ at the latest epoch.

Applying the preferred host attenuation of $A_V=3.3$~mag yields a markedly different and more physically plausible SED evolution. The peak luminosity increases to $\log(L_{\rm bb}/{\rm erg\,s^{-1}})\simeq43.8$, declines to $\simeq43.4$ by $\sim$+6~d, and reaches $\simeq40.8$ by $\sim$+27~d. The inferred radius is $R_{\rm bb}\simeq(2$--$3)\times10^{4}\,\Rsun$ 
around peak and contracts to $\sim1.4\times10^{4}\,\Rsun$ by $\sim$+6~d. The temperature correspondingly rises from $T_{\rm bb}\simeq7\times10^{3}$~K at $\sim$-4~d to $\simeq(1.3$--$1.5)\times10^{4}$~K near maximum, remains near $\simeq1.3$--$1.4\times10^{4}$~K through the first week, and cools to $\sim2\times10^{3}$~K at the latest epoch. These early-time temperatures are broadly consistent with the hot continuum typically associated with optical \ion{C}{2} in SN spectra. \textcolor{black}{The inferred blackbody radius of $R_{\rm bb}\simeq(2$--$3)\times10^{4}~R_\odot$ ($\simeq1.4\times10^{15}$--$2.1\times10^{15}$~cm; $\simeq93$--$140$~AU) 
near peak is inconsistent with a bare stellar photosphere and instead points to an optically thick, geometrically confined circumstellar envelope at scales of $\sim10^{2}$~AU, consistent with pre-SN mass loss or an LBV-like eruption occurring years to decades prior to explosion. The subsequent contraction to $R_{\rm bb}\sim1.4\times10^{4}~R_\odot$ ($\sim9.7\times10^{14}$~cm; $\sim65$~AU) by $\sim$+6~d, combined with 
the rising photospheric temperature toward maximum light, is consistent with the forward shock propagating through and progressively consuming the optically thick CSM, reducing the effective photospheric radius as 
the swept-up shell becomes increasingly optically thin. Crucially, at $t\sim+6$~d the blackbody radius remains significantly larger than the inferred ejecta radius of $R_{\rm ej}\simeq(1.66$--$1.97)\times10^{14}$~cm 
($\simeq11$--$13$~AU), confirming that the photosphere is still embedded within the overlying unshocked CSM at this epoch.}

Overall, the extinction-corrected fits are more consistent with the spectroscopic evidence for a hot radiation field, whereas the uncorrected fits imply temperatures that are too low to naturally explain the observed high-ionization features. 

\subsection{CSM-interaction lightcurve semi-analytical modeling}
\label{sec:lcmodelling}


We model the multi-band lightcurve using the semi-analytic circumstellar interaction framework of \citet{Chatzopoulos2012,Chatzopoulos2013}, as implemented in \texttt{redback} \citep{sarin_redback}. The model assumes that the observed luminosity is powered by conversion of kinetic energy into radiation at the forward and reverse shocks, followed by diffusion through an optically thick CSM. The CSM is described by a power-law density profile,
\begin{equation}
\rho_{\rm csm}(r) = q\,r^{-\eta}, \qquad q \equiv \rho_0\,r_0^{\eta},
\end{equation}
where $r_0$ is the inner CSM radius and $\rho_0\equiv \rho_{\rm csm}(r_0)$ is the density at $r_0$. For a given $M_{\rm csm}$, the outer edge of the CSM follows from mass conservation,
\begin{equation}
M_{\rm csm}=\int_{r_0}^{R_{\rm csm}}4\pi r^2 \rho_{\rm csm}(r)\,dr
= \frac{4\pi q}{3-\eta}\left(R_{\rm csm}^{3-\eta}-r_0^{3-\eta}\right),
\end{equation}

The corresponding optically thick CSM mass is
\begin{equation}
M_{\rm csm,th}=\frac{4\pi q}{3-\eta}\left(r_{\rm ph}^{3-\eta}-r_0^{3-\eta}\right),
\end{equation}
where $r_{\rm ph}$ is the photospheric radius within the CSM. Following \citet{Chatzopoulos2012,Chatzopoulos2013}, we adopt an outer ejecta density profile index $n=12$ \citep{Matzner1999} and inner slope $\delta=0$ (shell-like CSM structure). The shock-crossing time of the optically thick CSM, $t_{\rm sh}$, is computed using the self-similar shock solutions and tabulated constants of \citet{Chatzopoulos2012,Chatzopoulos2013}, and we define the diffusion time through the optically thick CSM as
\begin{equation}
t_{\rm diff}=\left(\frac{2\,\kappa\,M_{\rm csm,th}}{13.7\,c\,v_{\rm ej}}\right)^{1/2}.
\end{equation}
For each model evaluation, we convert the bolometric luminosity and photospheric radius into broad-band fluxes using a blackbody spectral energy distribution with a temperature floor,
\[
T=\max\!\left[\left(\frac{L}{4\pi\sigma r_{\rm ph}^2}\right)^{1/4},\,T_{\rm floor}\right],
\]
and include host extinction as a free parameter, $A_V^{\rm host}$. We fix the redshift to $z=0.0294$ and adopt Gaussian priors on the opacity parameters, centered at $\kappa=0.07~{\rm cm^2\,g^{-1}}$ and $\kappa_\gamma=0.03~{\rm cm^2\,g^{-1}}$, each with a width of $0.001$, effectively treating both opacities as constant. We use log-uniform priors on $M_{\rm ej}$, $M_{\rm csm}$, $v_{\rm ej}$, $\rho$, $r_0$, and $T_{\rm floor}$, and uniform priors on $\eta$, $t_0$, and $A_V^{\rm host}$. Specifically, the ejecta mass and CSM mass are allowed to vary over $0.1$--$15$~M$_\odot$ and $10^{-3}$--$15$~M$_\odot$, respectively, while the ejecta velocity spans $10^3$--$5\times10^4$~km~s$^{-1}$. The inner CSM radius is assigned a log-uniform prior over $0.1$--$100$~AU, and the density at the inner radius is given a log-uniform prior over $10^{-15}$--$10^{-9}$~g~cm$^{-3}$. The CSM density profile index is assigned a uniform prior between $\eta=0$, corresponding to a shell-like configuration, and $\eta=2$, corresponding to a steady wind. Posterior sampling is performed using nested sampling \citep{Feroz2009} with the \texttt{dynesty} package \citep{Speagle2020}, adopting the \texttt{rslice} sampling method. 

\begin{figure}
    \centering
    \includegraphics[width=\columnwidth]{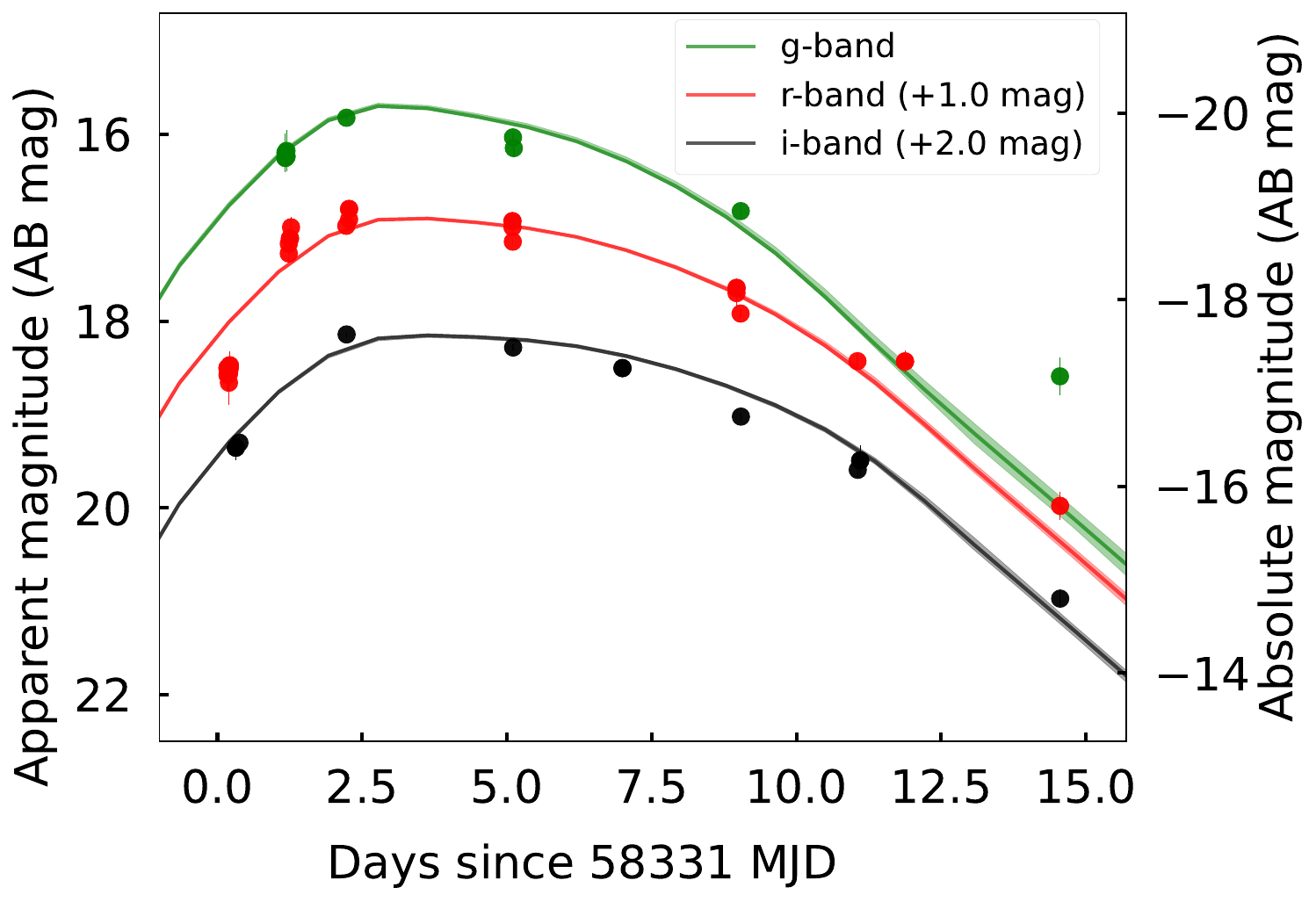}
    \caption{Multi-band light-curve fit of SN~2018erx using the CSM-interaction model implemented in \texttt{redback}. 
The observed photometry in the $g$, $r$, and $i$ bands is shown as colored points, while the solid curves represent the best-fitting model.}
    \label{fig:modelfit}
\end{figure}

\begin{figure*}
    \centering
    \includegraphics[width=17cm]{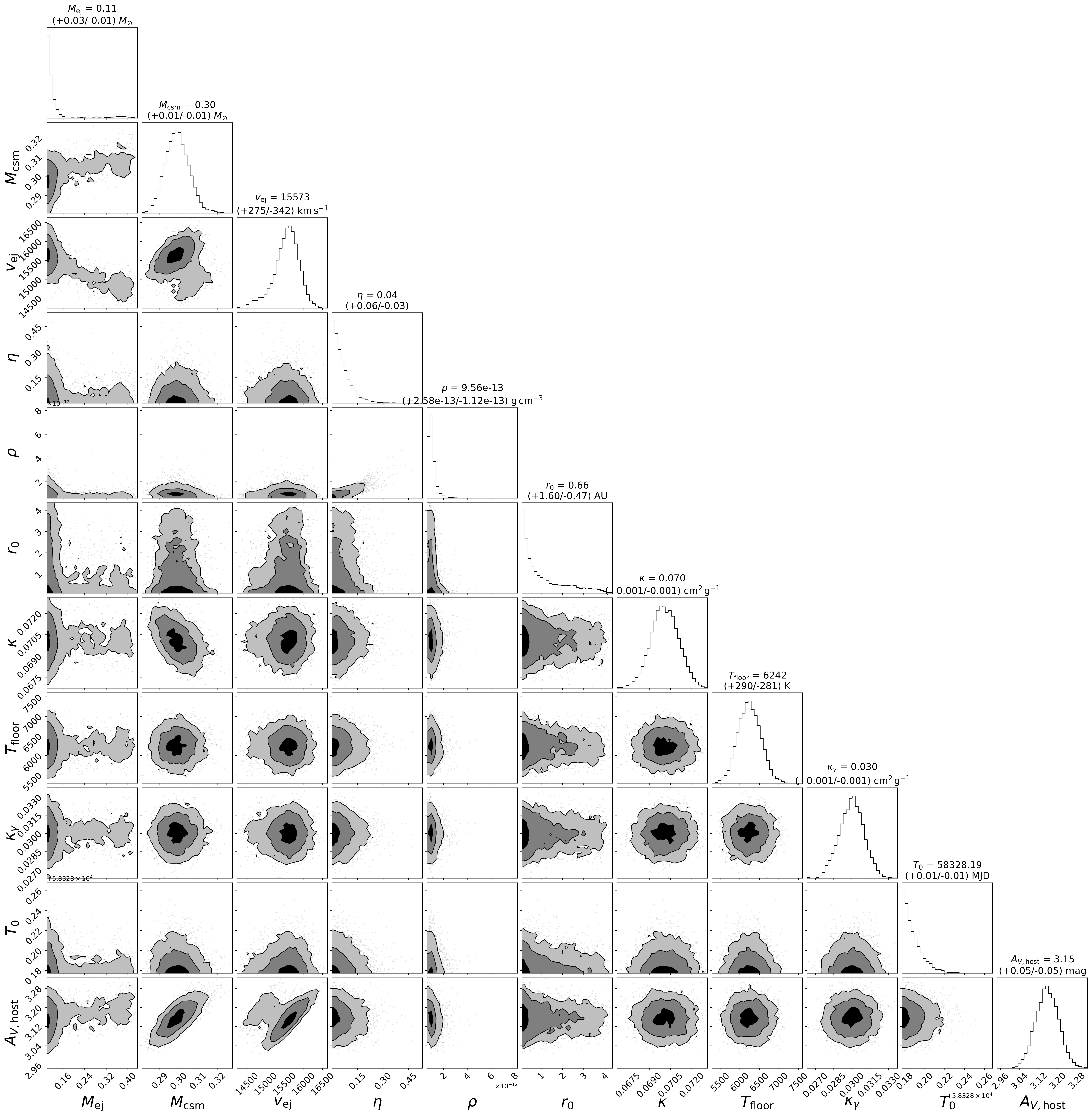}
    \caption{Corner plot showing the posterior distributions of the CSM-interaction model parameters obtained from the \texttt{redback} fit to the multi-band lightcurve of SN~2018erx. The diagonal panels show the marginalized posterior distributions, while the off-diagonal panels show the parameter covariances. The dashed lines indicate the median values and $1\sigma$ credible intervals. 
}
    \label{fig:corner}
\end{figure*}

The priors and the best-fit parameters are summarized in \autoref{tab:csm_priors_constraints} and \autoref{tab:csm_posterior}, respectively. The resulting multi-band lightcurve fit is shown in Figure~\ref{fig:modelfit}. The quoted 1$\sigma$ uncertainties reflect only the formal statistical errors derived from the fitting procedure and do not account for systematic or model-dependent uncertainties. We note that r$_{0}$ and $\rho_{0}$ are driven toward the boundaries of the explored parameter space (see Figure ~\ref{fig:corner}). 

We find that a CSM mass of
$M_{\rm CSM}\simeq 0.30^{+0.01}_{-0.01}\,{\rm M_\odot}$ provides a good reproduction of the observed lightcurves when coupled with an ejecta mass of
$M_{\rm ej}\simeq 0.11^{+0.03}_{-0.01}\,{\rm M_\odot}$ and an interaction onset at an inner CSM radius of
$R_0 \equiv r_0 \simeq 0.66^{+1.60}_{-0.47}\,\mathrm{AU}$.
The fit further prefers substantial host extinction,
$A_V \simeq 3.15^{+0.05}_{-0.05}$~mag.
The comparatively low ejecta mass is broadly consistent with expectations for ultra-stripped explosions from low-mass helium stars in close binary systems \citep{Das2024,Wu2022}.

\begin{figure*}
    \centering
    \includegraphics[width=9.5cm]{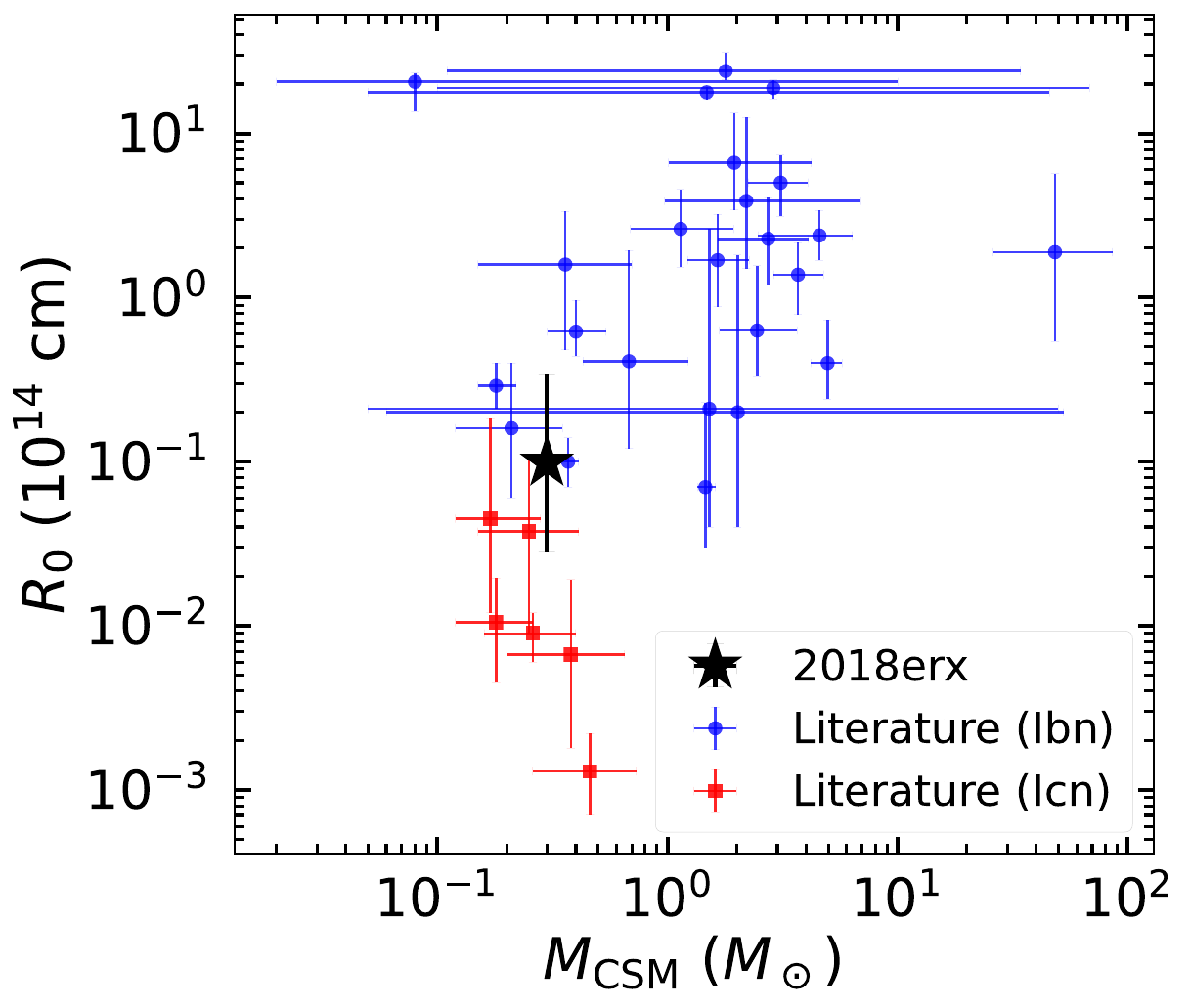}
    \includegraphics[width=12.5cm]{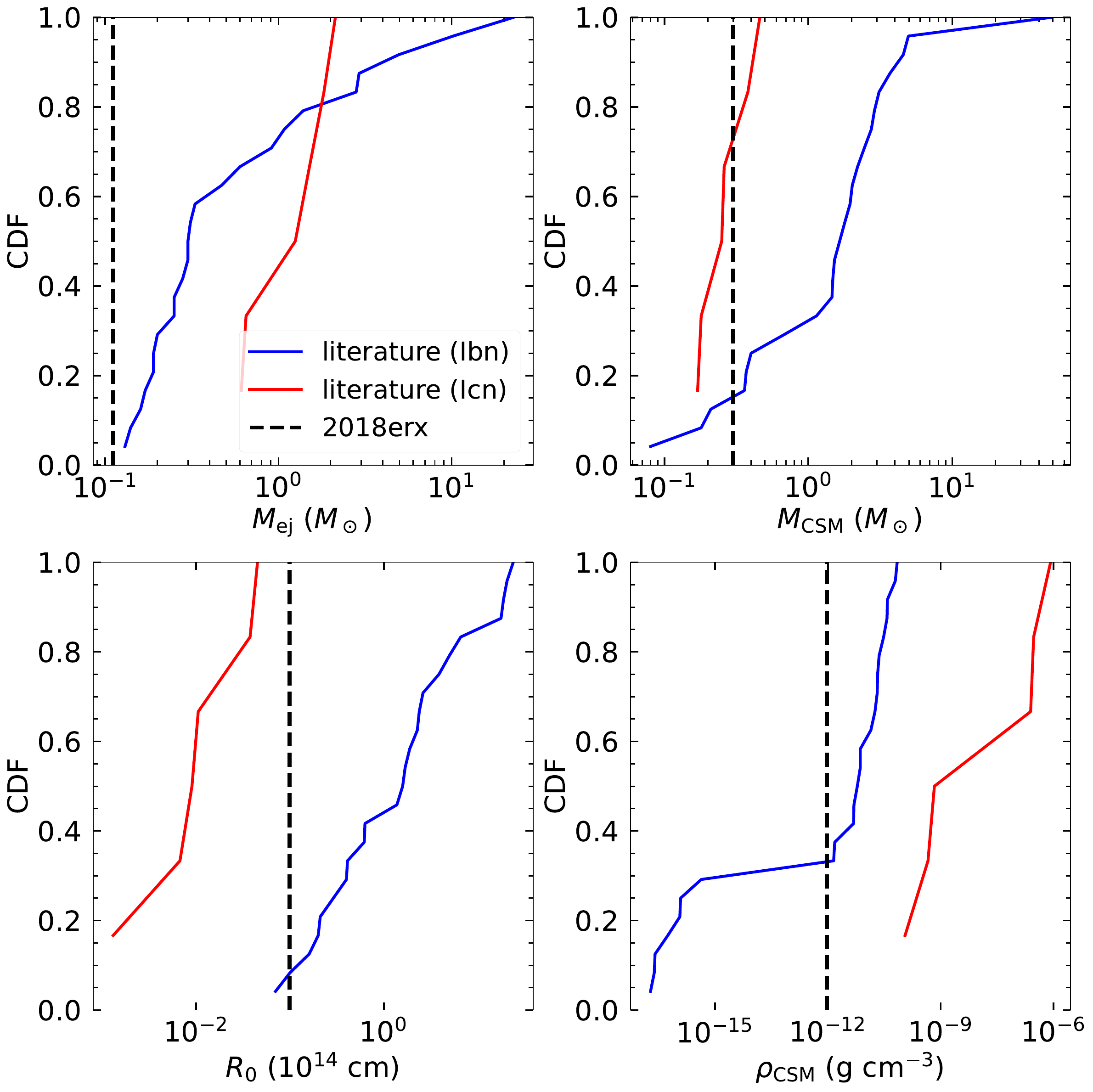}
\caption{Comparison of the circumstellar interaction parameters inferred for SN~2018erx with literature samples of Type~Ibn/Icn SNe. Left: Inner CSM radius $R_0$ versus CSM mass $M_{\rm CSM}$. The black star marks the values inferred for SN~2018erx from the \texttt{redback} interaction modeling, while blue and red symbols show literature estimates for SNe~Ibn and SNe~Icn, respectively. Right: Cumulative distribution functions (CDFs) of the inferred interaction parameters ($M_{\rm ej}$, $M_{\rm CSM}$, $R_0$, $v_{\rm ej}$, and $\rho_{\rm CSM}$) for literature samples of SNe~Ibn (blue) and SNe~Icn (green). The dashed vertical line indicates the corresponding value for SN~2018erx.}
\label{fig:csmliterature}
\end{figure*}

\textcolor{black}{Compared with literature interaction-modeling results for SNe~Ibn and SNe~Icn \citep[e.g.,][]{Pellegrino2022b,Farias2025}, SN~2018erx occupies the low-mass end of both the ejecta-mass and CSM-mass distributions (Figure~\ref{fig:csmliterature}). In the $R_0$--$M_{\rm CSM}$ plane, it 
lies well below the bulk of the SN Ibn population, indicating a more compact inner CSM radius at comparable CSM mass, while remaining broadly consistent with the compact radii inferred for the small number of modeled SN Icn events. The cumulative-distribution comparison likewise shows that $M_{\rm ej}$, $M_{\rm CSM}$, and CSM density for SN~2018erx fall toward the lower end of the Type Ibn distribution, yet remain within the range spanned by both the Type Ibn and Icn samples. The CSM model fitted ejecta velocity, $v_{\rm ej}\simeq 1.5\times10^{4}$~km~s$^{-1}$, lies toward the higher end of the SN Ibn 
distribution and is comparable to values inferred for SNe~Icn, suggesting a faster-moving, lower-mass ejecta component more reminiscent of stripped-envelope explosions with little residual helium.}

\textcolor{black}{The inferred CSM density profile, characterized by a shallow power-law  slope $\eta \simeq 0.05^{+0.06}_{-0.32}$, is the most distinctive feature of SN~2018erx within the broader Type Ibn/Icn population. This value is indicative of a shell-like, approximately constant-density CSM 
configuration, in stark contrast to the $\eta = 2$ profile expected for a steady stellar wind. While \citet{Farias2025} report that a subset of SNe~Ibn exhibit shell-like CSM geometries, the majority of events in their sample are better described by wind-like density profiles with 
$\eta \sim 2$. Among SNe~Icn, the modeled sample remains small, but the available results suggest similarly compact and dense CSM configurations, with some events also preferring shallower density slopes \citep{Pellegrino2022b}. 
}

\textcolor{black}{Taken together, the combination of small $R_0$, high CSM density, low ejecta mass, high ejecta velocity, and a nearly flat density profile disfavors a canonical steady-state WR wind as the dominant pre-explosion mass-loss mechanism. Instead, these properties are more naturally explained by a brief, intense episode of mass ejection in the years to decades prior to explosion, forming a confined, shell-like CSM. Such behavior is more consistent with binary-driven mass transfer or common-envelope interaction than with the continuous radiatively-driven outflows associated with isolated WR stars. We return to this point in Section~\ref{sec:progenitordiscussion}, where we argue  that the full set of observational and modeling constraints for SN~2018erx collectively favor a binary progenitor channel over a single 
massive-star scenario.}

To relate the fitted parameters to an approximate CSM outer radius, we adopt a power-law density profile
\begin{equation}
\rho(r) = \rho_0 \left(\frac{r}{R_0}\right)^{-\eta}, \qquad R_0 \le r \le R_{\rm out},
\end{equation}
where $\rho_0$ is the density at the inner edge ($\rho_0 \simeq 9.56^{+2.50}_{-1.12}\times10^{-13}\,\mathrm{g\,cm^{-3}}$).
The total CSM mass then satisfies
\begin{equation}
M_{\rm CSM} = 4\pi \int_{R_0}^{R_{\rm out}} \rho(r)\,r^2\,dr
= \frac{4\pi \rho_0 R_0^{3}}{3-\eta}
\left[\left(\frac{R_{\rm out}}{R_0}\right)^{3-\eta}-1\right],
\label{eq:Mcsm_powerlaw}
\end{equation}
for $\eta \neq 3$. Inverting Eq.~(\ref{eq:Mcsm_powerlaw}) gives
\begin{equation}
R_{\rm out} = R_0\left[1+\frac{(3-\eta)\,M_{\rm CSM}}{4\pi \rho_0 R_0^{3}}\right]^{1/(3-\eta)}.
\end{equation}

Using the posterior median values, we obtain
\begin{equation}
R_{\rm out} \simeq 5.6\times10^{14}\ {\rm cm} \simeq 37~{\rm AU}.
\end{equation}
The inferred CSM therefore extends from $R_0\sim0.6$~AU to $\sim40$~AU.



\begin{table}
\centering
\caption{Priors and constraints adopted for the \texttt{redback} CSM interaction fit.}
\label{tab:csm_priors_constraints}
\begin{tabular}{l c c}
\hline
Parameter & Prior type & Range / Value \\
\hline
$z$ & fixed & $0.0294$ \\
$M_{\mathrm{ej}}$ [M$_\odot$] & log-uniform & $[0.1,\,15]$ \\
$v_{\mathrm{ej}}$ [km\,s$^{-1}$] & log-uniform & $[10^3,\,5\times10^4]$ \\
$M_{\mathrm{CSM}}$ [M$_\odot$] & log-uniform & $[10^{-3},\,15]$ \\
$\eta$ & uniform & $[0,\,2]$ \\
$\rho$ [g\,cm$^{-3}$] & log-uniform & $[10^{-15},\,10^{-9}]$ \\
$r_0$ [AU] & log-uniform & $[10^{-1},\,10^{2}]$ \\
$T_{\mathrm{floor}}$ [K] & log-uniform & $[10^{2},\,10^{4}]$ \\
$t_0$ [MJD] & uniform & $[58321.18,\,58331.18]$ \\
$A_V$ [mag] & uniform & $[0,\,5]$ \\
$\kappa$ [cm$^{2}$\,g$^{-1}$] & Gaussian & $0.07 \pm 0.001$ \\
$\kappa_\gamma$ [cm$^{2}$\,g$^{-1}$] & Gaussian & $0.03 \pm 0.001$ \\
$t_{\mathrm{diff}}/t_{\mathrm{shock}}$ & constraint & $[0.1,\,0.8]$ \\
$r_{\mathrm{ph}}/r_{\mathrm{CSM}}$ & constraint & $[0,\,1]$ \\
$r_{0}/r_{\mathrm{ph}}$ & constraint & $[0,\,1]$ \\

\hline
\end{tabular}
\end{table}

\begin{table}
\centering
\caption{Posterior median values and 68\% credible intervals (16th and 84th percentiles) for the CSM interaction model parameters.}
\label{tab:csm_posterior}
\begin{tabular}{l c}
\hline
Parameter & Value \\
\hline

$M_{\mathrm{ej}}$ [M$_\odot$] 
& $0.111^{+0.034}_{-0.009}$ \\

$M_{\mathrm{CSM}}$ [M$_\odot$] 
& $0.299^{+0.006}_{-0.006}$ \\

$v_{\mathrm{ej}}$ [km\,s$^{-1}$] 
& $(1.56^{+0.03}_{-0.03}) \times 10^{4}$ \\

$\eta$ 
& $0.045^{+0.064}_{-0.033}$ \\

$\rho$ [g\,cm$^{-3}$] 
& $(9.56^{+2.58}_{-1.12}) \times 10^{-13}$ \\

$r_0$ [AU] 
& $0.66^{+1.60}_{-0.47}$ \\

$T_{\mathrm{floor}}$ [K] 
& $6242^{+290}_{-281}$ \\

$t_0$ [MJD] 
& $58328.19^{+0.01}_{-0.01}$ \\

$A_V$ [mag] 
& $3.15^{+0.05}_{-0.05}$ \\

\hline
\end{tabular}
\end{table}

\subsection{CSM radial extent based on spectroscopy}

\label{sec:csm_radius_spec}

The spectra of SN~2018erx show prominent intermediate-width \ion{C}{2} emission and related interaction signatures (Figures~\ref{fig:spec_all} and \ref{fig:CII}), indicating that the line-forming region is associated with shocked circumstellar gas. These features can therefore be used to place order-of-magnitude constraints on the radial extent of the CSM. We detect clear interaction signatures already in the first spectrum at MJD~58334, corresponding to roughly 6~d after the explosion epoch inferred in Section~\ref{sec:explosion}. This implies that the fastest interaction front, identified with the outer ejecta or the forward shock, must have reached the inner edge of the CSM by that time. 

For this estimate, we adopt a characteristic ejecta velocity of $v_{\rm ej}\simeq5200~{\rm km~s^{-1}}$ from the P-Cygni absorption minima. This gives
\begin{equation}
R_{\rm CSM,\,inner} \lesssim v_{\rm ej}\,\Delta t \simeq 18~{\rm AU},
\end{equation}
consistent with an extended circumstellar envelope at the time of interaction. Interaction-related \ion{C}{2} absorption/emission remains visible as late as MJD~58351, about 23~d after explosion. This indicates that CSM-processed material continues to contribute to the line-forming region at least until that epoch. Using the same velocity then gives a conservative lower limit on the outer extent of the line-forming CSM,
\begin{equation}
R_{\rm CSM,\,outer} \gtrsim v_{\rm ej}\,\Delta t \simeq 70~{\rm AU}.
\end{equation}
These spectroscopic constraints are broadly consistent with the lightcurve modeling. The interaction fit yields an inner CSM radius of $R_0 \simeq 0.56$~AU, comfortably below the spectroscopic upper limit on the inner edge, while the outer radius inferred from the best-fit CSM parameters is $R_{\rm out}\sim40$~AU, comparable to the spectroscopic lower limit above.

\subsection{Order-of-magnitude CSM mass estimate from ejecta--CSM interaction}
\label{sec:csm_mass_oome}
We provide an independent, order-of-magnitude estimate of the CSM mass by treating the early emission as powered by kinetic energy dissipated in a strongly inelastic ejecta--CSM collision. Here we explicitly distinguish the \emph{pre-shock} CSM speed ($v_{\rm w}$) from the \emph{post-shock} bulk speed of the swept-up shell ($v_{\rm shell}$) inferred from the absorption.

Consider an outer ejecta layer of mass $M_{{\rm ej},i}$ moving at $v_{\rm ej}$ that impacts a CSM shell of mass $M_{\rm CSM}$ moving at a pre-shock speed $v_{\rm w}$. In the perfectly inelastic limit, the shocked ejecta and swept-up CSM merge into a single shell (e.g. a cooled dense shell) moving at a common post-shock velocity $v_{\rm shell}$. Conservation of momentum gives
\begin{equation}
(M_{{\rm ej},i}+M_{\rm CSM})\,v_{\rm shell}
= M_{{\rm ej},i}\,v_{\rm ej} + M_{\rm CSM}\,v_{\rm w},
\label{eq:mom_inelastic}
\end{equation}
or equivalently
\begin{equation}
v_{\rm shell}-v_{\rm w}
= \frac{M_{{\rm ej},i}}{M_{{\rm ej},i}+M_{\rm CSM}}\,(v_{\rm ej}-v_{\rm w}).
\label{eq:vsh_relation}
\end{equation}

The kinetic energy dissipated in the collision is the difference between the initial and final kinetic energies,
\begin{align}
\Delta E
&= \frac{1}{2}\,M_{{\rm ej},i}\,v_{\rm ej}^{2}
 + \frac{1}{2}\,M_{\rm CSM}\,v_{\rm w}^{2}
 - \frac{1}{2}\,(M_{{\rm ej},i}+M_{\rm CSM})\,v_{\rm shell}^{2} \nonumber\\
&= \frac{1}{2}\,\frac{M_{{\rm ej},i}\,M_{\rm CSM}}{M_{{\rm ej},i}+M_{\rm CSM}}\,(v_{\rm ej}-v_{\rm w})^{2}.
\label{eq:DeltaE_general}
\end{align}
Using Eq.~\eqref{eq:vsh_relation} to eliminate the mass ratio yields a convenient expression in terms of the observable post-shock velocity:
\begin{equation}
\Delta E
= \frac{1}{2}\,M_{\rm CSM}\,(v_{\rm ej}-v_{\rm w})\,(v_{\rm shell}-v_{\rm w}).
\label{eq:DeltaE_vsh}
\end{equation}
For the order-of-magnitude estimate we adopt $v_{\rm w}\ll v_{\rm shell}$, so that
\begin{equation}
\Delta E \simeq \frac{1}{2}\,M_{\rm CSM}\,v_{\rm ej}\,v_{\rm shell}.
\label{eq:DeltaE_vsh_simplified}
\end{equation}

If the interaction region is radiative on the relevant timescale, a fraction $\epsilon \le 1$ of the dissipated energy emerges as radiation, $E_{\rm rad}\simeq \epsilon\,\Delta E$. Solving Eq.~\eqref{eq:DeltaE_vsh} for the CSM mass gives
\begin{equation}
M_{\rm CSM}
\simeq
\frac{2\,E_{\rm rad}}{\epsilon\,(v_{\rm ej}-v_{\rm w})\,(v_{\rm shell}-v_{\rm w})}
\;\;\approx\;\;
\frac{2\,E_{\rm rad}}{\epsilon\,v_{\rm ej}\,v_{\rm shell}}
\quad (v_{\rm w}\ll v_{\rm shell}).
\label{eq:Mcsm_from_Erad_vsh}
\end{equation}



Adopting $v_{\rm ej}\simeq 5200~{\rm km~s^{-1}}$ from the P-Cygni absorption minima, we estimate the post-shock velocity from the \ion{C}{2} $\lambda7238$ emission feature. After fitting a local continuum within $\pm50$~\AA\ of line center and modeling the line profile with Gaussian and Lorentzian functions, we infer a characteristic velocity of $v_{\rm shell}=3900\pm100~{\rm km~s^{-1}}$,
which we interpret as the bulk speed of the post-shock swept-up shell. We thus obtain

\begin{equation}
\label{eq:Mcsm_num_vsh_50}
\begin{aligned}
M_{\rm CSM}\simeq\;&
4.9\times10^{-1}\,
\epsilon^{-1}\,
\left(\frac{E_{\rm rad}}{10^{50}\ {\rm erg}}\right)
\left(\frac{v_{\rm ej}}{5200\ {\rm km~s^{-1}}}\right)^{-1}\\
&\times
\left(\frac{v_{\rm shell}}{3900\ {\rm km~s^{-1}}}\right)^{-1}
\ {\rm M_\odot}.
\end{aligned}
\end{equation}

The inferred deceleration also fixes the interacting mass ratio. From Eq.~\eqref{eq:mom_inelastic}, for $v_{\rm w}\ll v_{\rm shell}$ one finds
\begin{equation}
\frac{M_{\rm CSM}}{M_{{\rm ej},i}}
\simeq
\left(\frac{v_{\rm ej}}{v_{\rm shell}}\right)-1
\approx 0.33,
\label{eq:massratio_from_vsh}
\end{equation}
so the swept-up CSM mass is comparable to the interacting ejecta-layer mass for the adopted velocities, consistent with the regime of efficient dissipation. This estimate should be interpreted as an order-of-magnitude scale. If the interaction is not fully radiative on the observed timescale, $\epsilon<1$ increases the inferred $M_{\rm CSM}$. Thus, for reasonable choices of $\epsilon$ and $E_{\rm rad}$, this order-of-magnitude estimate is broadly consistent with the CSM mass inferred from the \texttt{redback} interaction fit ($M_{\rm CSM}\simeq0.31\,M_\odot$), given the large systematic uncertainties inherent in the simplified inelastic-collision picture.

\subsection{Mass-loss-rate estimates for the inner CSM}
\label{sec:mdot_inner_csm}


Below we estimate the characteristic mass-loss scale associated with the inner CSM: one based on the observed luminosity and spectroscopic line width, and a second based on the the best-fit CSM model.

\subsubsection{Based on spectroscopy}

If the observed luminosity is powered primarily by kinetic-energy dissipation at the ejecta--CSM interaction shock, the corresponding mass-loss rate can be estimated as
\begin{equation}
\dot{M} = \frac{2 L v_{w}}{\epsilon\, v_{\mathrm{shell}}^{3}},
\end{equation}
where $\epsilon \leq 1$ is the efficiency with which shock kinetic energy is converted into radiation, $v_w$ is the pre-SN outflow velocity, $v_{\rm shell}$  is the velocity of the post-shock shell, and $L$ is the bolometric luminosity. 

The outflow velocity is not directly measured in SN~2018erx, so we consider two representative values: $v_w=100~{\rm km~s^{-1}}$, appropriate for a slower eruptive or binary-driven outflow, and $v_w=1000~{\rm km~s^{-1}}$, characteristic of a WR-like wind. To estimate the post-shock velocity, we analyze the \ion{C}{2} $\lambda7238$ emission feature. After fitting a local continuum within $\pm50$~\AA\ of line center, we model the profile with Gaussian and Lorentzian functions and infer a characteristic velocity of $v_{\rm shell}=3900\pm100~{\rm km~s^{-1}}.$

Adopting $\epsilon=0.5$, we obtain characteristic mass-loss rates of
\[
\dot{M} \approx 4.7\times10^{-4}\,M_\odot\,{\rm yr^{-1}}
\]
for $v_w=100~{\rm km~s^{-1}}$, and
\[
\dot{M} \approx 4.7\times10^{-3}\,M_\odot\,{\rm yr^{-1}}
\]
for $v_w=1000~{\rm km~s^{-1}}$.

\subsubsection{Effective mass-loss rate from the CSM model fitting}

Although the best-fit density profile is shell-like rather than a steady wind, it is still useful to quote an ``effective'' mass-loss rate by identifying the fitted density normalization $\rho_0$ at $R_0$ with that of a stationary wind:
\begin{equation}
\dot{M}_{\rm eff} = 4\pi R_0^2 \rho_0 v_w.
\end{equation}
Using the posterior median values of $R_0$ and $\rho_0$, and adopting a fiducial WR-like velocity of $v_w=1000~{\rm km~s^{-1}}$, we find
\[
\dot{M}_{\rm eff} \sim 1.9\times10^{-3}\,M_\odot\,{\rm yr^{-1}}.
\]
This should be interpreted only as a characteristic rate at the inner CSM radius, not as evidence for a long-lived steady wind.

The spectroscopic and model-based estimates are broadly consistent at the order-of-magnitude level. For $v_w=1000~{\rm km~s^{-1}}$, the luminosity-based estimate exceeds the effective rate inferred from the CSM fit by a factor of a few; given the different assumptions involved, this level of agreement is reasonable. Both approaches therefore indicate a dense and enhanced pre-SN mass-loss episode.


\begin{figure*}
    \centering
    \includegraphics[width=14.5cm]{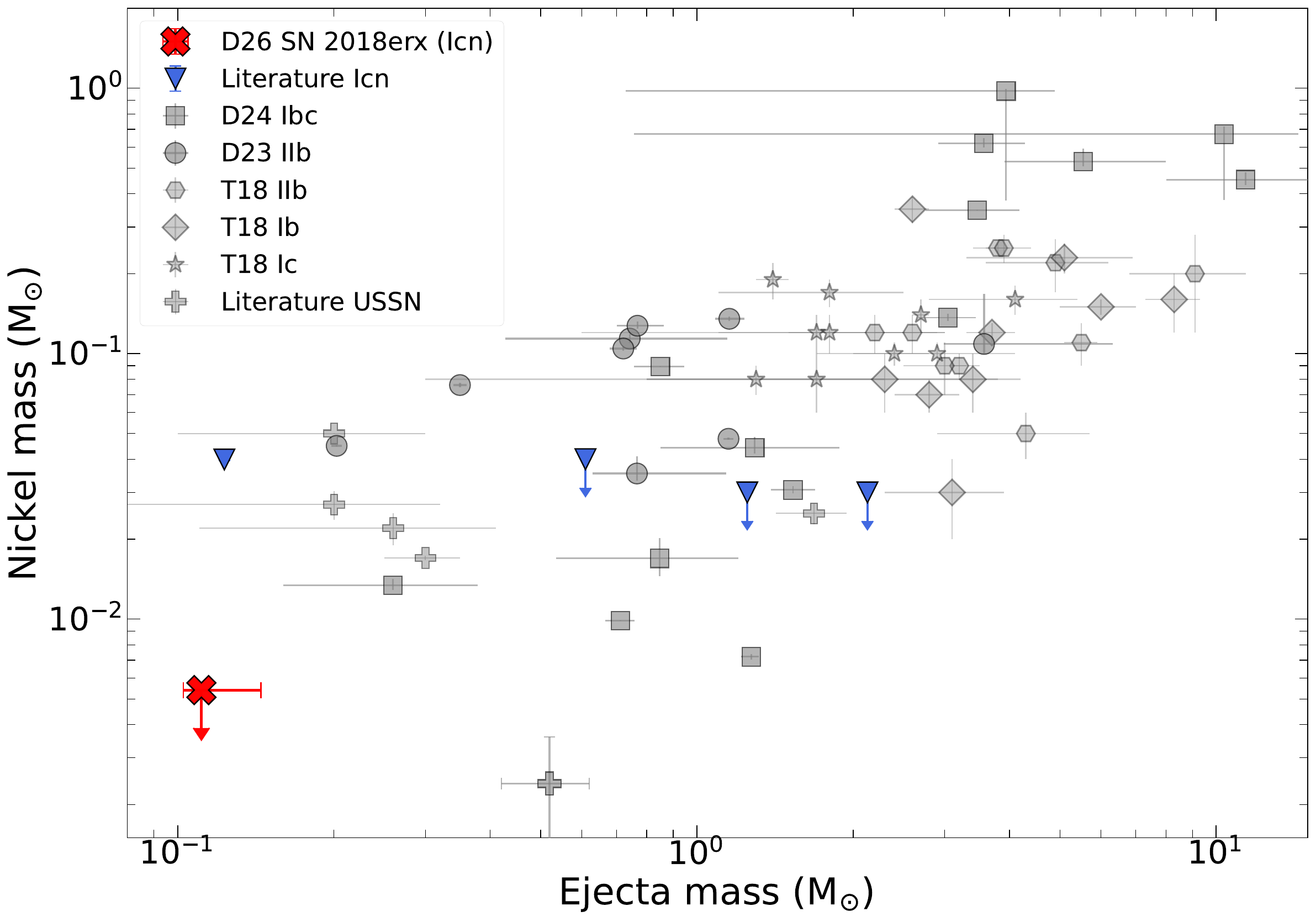}
\caption{
Comparison of the $^{56}$Ni mass ($M_{\rm Ni}$) and ejecta mass ($M_{\rm ej}$) of SN~2018erx with SESNe from the literature. 
The red symbol marks SN~2018erx, while blue triangles denote Type~Icn events \citep{Pellegrino2022b}. 
The arrow indicates that the inferred $^{56}$Ni mass is an upper limit. 
Gray symbols show literature measurements for other SESN subclasses from D24 \citep{Das2023b}, D23 \citep{Das2023a}, and T18 \citep{Taddia2018}, as well as ultra-stripped SN candidates (USSN) from \citet{Das2024}. 
SN~2018erx occupies the regime of both low ejecta mass and low $^{56}$Ni mass.
}
    \label{fig:mnimej}
\end{figure*}

\subsection{Upper limits on the $^{56}$Ni mass in SN~2018erx from the rapidly fading lightcurve}
\label{sec:ni_mass}

At late phases, the luminosity of core-collapse SNe is often powered by radioactive decay. In events with a well-detected nebular tail, the $^{56}$Ni mass can be inferred by comparing the late-time bolometric luminosity to the expected $^{56}$Co decay power. For SN~2018erx, however, the available photometry does not reveal a clear $^{56}$Co-powered tail. We therefore treat the luminosity measured at the last blackbody epoch as an upper limit on the contribution from radioactive decay and derive corresponding upper limits on $M_{\mathrm{Ni}}$.

For an initial $^{56}$Ni mass $M_{\rm Ni}$, the instantaneous radioactive
power from the full $^{56}$Ni $\rightarrow$ $^{56}$Co $\rightarrow$
$^{56}$Fe decay chain is
\begin{equation}
L_{\rm rad}(t)=
f_{\rm dep}(t)
\left(\frac{M_{\rm Ni}}{M_\odot}\right)
\epsilon_{\rm rad}(t),
\label{eq:lrad}
\end{equation}
where
\begin{equation}
\begin{split}
\epsilon_{\rm rad}(t) ={}&
6.45\times10^{43} e^{-t/8.8\,{\rm d}}  \\
&+1.45\times10^{43} e^{-t/111.3\,{\rm d}}
\quad {\rm erg~s^{-1}} .
\end{split}
\label{eq:eps_rad}
\end{equation}
Here $t$ is the rest-frame time since explosion, the first term represents
heating from $^{56}$Ni decay, and the second term represents heating from
$^{56}$Co decay. The factor $f_{\rm dep}(t)$ accounts for incomplete
deposition of the radioactive energy; for complete trapping,
$f_{\rm dep}=1$.

Requiring $L_{\rm rad}(t_{\rm last}) \leq L_{\rm bb}(t_{\rm last})$ gives
\begin{equation}
M_{\rm Ni} \leq
\frac{L_{\rm bb}(t_{\rm last})}
{f_{\rm dep}(t_{\rm last})\,
\epsilon_{\rm rad}(t_{\rm last})}
M_\odot .
\label{eq:mni_ul_fullchain}
\end{equation}

The last blackbody epoch that passes our quality cut occurs at phase
$+27.7$~d relative to $r$-band maximum, corresponding to
$t_{\rm last}\simeq34$~d after explosion. Assuming complete trapping, the no-host-extinction luminosity at this epoch,
$\log L_{\rm bb}({\rm erg~s^{-1}})=40.63$, gives
\begin{equation}
M_{\rm Ni} \lesssim 3.5\times10^{-3}~M_\odot .
\end{equation}
For the high-extinction case, the
corresponding limit is
\begin{equation}
M_{\rm Ni} \lesssim 5.4\times10^{-3}~M_\odot .
\end{equation}

These limits should be interpreted cautiously, since the last blackbody epoch is likely dominated at by
hot dust or reprocessed emission rather than a clean nebular radioactive tail as discussed in Section \ref{sec:dust}.
We therefore write the more general limit as
\begin{equation}
M_{\rm Ni} \lesssim
\left(\frac{f_{\rm rad}}{f_{\rm dep}}\right)
M_{\rm Ni,trap},
\label{eq:mni_factors}
\end{equation}
where $M_{\rm Ni,trap}$ is the complete-trapping value above, $f_{\rm rad}\leq1$
is the fraction of the observed late-time luminosity actually powered by
radioactive decay, and $f_{\rm dep}\leq1$ is the radioactive-energy deposition
fraction. Incomplete $\gamma$-ray trapping increases the inferred upper limit by
a factor $f_{\rm dep}^{-1}$, whereas any contribution from circumstellar
interaction, an infrared echo, or hot dust reprocessing reduces the radioactive
fraction of the luminosity and therefore lowers the allowed $^{56}$Ni mass. For ordinary SESN-like values $T_0\sim50$--150~d,
$f_{\rm dep}(34~{\rm d})\simeq0.89$--1.00, so the complete-trapping limits
remain nearly unchanged if $f_{\rm rad}=1$. Even in a more weakly trapped,
low-ejecta-mass configuration with $f_{\rm dep}\sim0.5$, the inferred limit
would increase only by a factor of $\sim2$. Thus, the two correction factors may partially compensate
each other, and the inferred $^{56}$Ni limits are likely to remain close to the
complete-trapping estimates above, i.e.
$M_{\rm Ni}\sim{\rm few}\times10^{-3}~M_\odot$.


\subsection{Dust}
\subsubsection{Dust Properties from the Near-Infrared Excess}
\label{sec:dust}

SN~2018erx exhibits a clear NIR excess at +27~d in addition to its unusually red optical continuum. Because the early-time spectra simultaneously show high-ionization carbon features (requiring a hot radiation field), 
the red continuum is difficult to reconcile with a genuinely cool photosphere alone, and instead naturally points to dust absorption and re-emission. In interacting SESNe, dust signatures can originate from (i) pre-existing circumstellar dust that survives the SN flash and produces an infrared echo, and/or (ii) newly formed dust condensing rapidly in the dense, cooling post-shock region (the cool dense shell; CDS).  Both pathways have been invoked for interacting SESNe, most famously in the Type~Ibn SN~2006jc \citep{Foley2007, Anupama2009}, and more recently in SN~2023xgo \citep{Gangopadhyay2025,Yamanaka2025}. Below we use the NIR excess to obtain order-of-magnitude constraints on the characteristic dust temperature, an effective emitting size, 
and the dust mass.

At the epoch of our NIR detection, the measured absolute magnitudes are $M_K=-14.25$ and
$M_J=-12.83$, giving $(J-K)_{\rm AB}=1.42$ mag and therefore
\begin{equation}
\frac{F_{\nu,J}}{F_{\nu,K}} = 10^{-0.4(J-K)_{\rm AB}} \simeq 0.270.
\label{eq:jk_fluxratio}
\end{equation}
If the emission is dominated by optically thin thermal dust emission, then
\begin{equation}
F_\nu \propto \kappa_\nu\,B_\nu(T_{\rm d}),
\end{equation}
where $\kappa_\nu$ is the dust mass absorption coefficient at frequency $\nu$. The observed color then constrains the dust temperature through
\begin{equation}
\frac{F_{\nu,J}}{F_{\nu,K}} =
\frac{\kappa_J}{\kappa_K}\,
\frac{B_\nu(\nu_J,T_{\rm d})}{B_\nu(\nu_K,T_{\rm d})}.
\label{eq:Tdust_ratio}
\end{equation}
For an order-of-magnitude estimate we take $\kappa_J/\kappa_K\simeq 1$ (a weak dependence in the
near-IR for typical grain models), which yields a color temperature
\begin{equation}
T_{\rm d} \simeq 1.66\times10^3~{\rm K}.
\end{equation}
We adopt a conservative range $T_{\rm d}\sim(1.5$--$2.0)\times10^3$~K for the hot component traced by
$J$ and $K$. This temperature is close to the condensation/sublimation temperature of carbonaceous
grains, consistent with a carbon-rich circumstellar environment as implied by the Type~Icn-like
spectral signatures.

If the NIR excess arises from an IR echo (i.e., dust at radii comparable to or larger than the
sublimation scale; see below), the emitting dust is expected to be optically thin in the near-IR for
the small dust masses implied by the data.  For a dust mass $M_{\rm d}$ distributed over a shell of
radius $R_{\rm d}$ and covering fraction $f_\Omega$, the characteristic K-band optical depth is
$\tau_K \sim \kappa_K M_{\rm d}/(4\pi f_\Omega R_{\rm d}^2)$, which is $\ll 1$ for
$R_{\rm d}\gtrsim{\rm few}\times10^{16}$~cm unless the dust is extremely clumped or confined to a very
small solid angle.

In the optically thin limit, the monochromatic luminosity is
\begin{equation}
L_{\nu,K} \simeq 4\pi\,\kappa_K\,M_{\rm d}\,B_\nu(\nu_K,T_{\rm d}),
\end{equation}
where $\kappa_K$ is the dust mass absorption coefficient at $2.2~\mu$m.
For spherical grains of radius $a$ and bulk density $\rho_{\rm b}$,
\begin{equation}
\kappa_\nu(a) = \frac{3\,Q_{\rm abs}(\nu,a)}{4a\rho_{\rm b}},
\label{eq:kappa_from_Q}
\end{equation}
where $Q_{\rm abs}(\nu,a)$ is the dimensionless absorption efficiency, i.e., the ratio of the absorption cross section to the geometric cross section of a grain. The above expression is equivalent to the commonly used grain-based form
\begin{equation}
M_{\rm d} =
\frac{4}{3}\,
\frac{a\rho_{\rm b}}{Q_{\rm abs}(\nu,a)}\,
\frac{F_\nu d^2}{B_\nu(\nu,T_{\rm d})},
\label{eq:mdust_Qform}
\end{equation}
where $F_\nu$ is the observed flux density at distance $d$.
For representative carbonaceous grains ($a\sim0.1~\mu{\rm m}$, $\rho_{\rm b}\sim2.2~{\rm g~cm^{-3}}$),
$Q_{\rm abs}$ of order a few $\times0.1$ at $\sim$2~$\mu$m corresponds to
$\kappa_K \sim 10^{4}~{\rm cm^2~g^{-1}}$, which we adopt as a fiducial value.

Solving for the dust mass gives
\begin{equation}
M_{\rm d} \simeq \frac{L_{\nu,K}}{4\pi\,\kappa_K\,B_\nu(\nu_K,T_{\rm d})}.
\label{eq:mdust}
\end{equation}
Adopting $\kappa_K = 10^4~{\rm cm^2~g^{-1}}$ yields
\begin{equation}
M_{\rm d} \simeq 1.2\times 10^{-6}~M_{\odot}\,
\left(\frac{\kappa_K}{10^4~{\rm cm^2~g^{-1}}}\right)^{-1},
\end{equation}
for $T_{\rm d}=1.66\times10^3$~K. The dominant systematic uncertainty is the poorly known near-IR
opacity, together with the uncertainty in $T_{\rm d}$ from the two-band color. Allowing
$T_{\rm d}=1500\text{--}2000$~K and $\kappa_K=10^3\text{--}10^4~{\rm cm^2~g^{-1}}$, we obtain
\begin{equation}
M_{\rm d} \sim 6\times 10^{-7}\text{--}2\times 10^{-5}~M_{\odot}
\end{equation}
for the hot dust component traced by the $J$- and $K$-band photometry.

For grains in radiative equilibrium with a central source of luminosity $L$, the
distance at which dust reaches temperature $T_{\rm d}$ is approximately

\begin{equation}
R(T_{\rm d}) \simeq
\left[
\frac{L}{16\pi\sigma T_{\rm d}^{4}}
\frac{\langle Q_{\rm abs}\rangle_{\rm in}}{\langle Q_{\rm abs}\rangle_{\rm out}}
\right]^{1/2},
\label{eq:Req}
\end{equation}

where $\langle Q_{\rm abs}\rangle_{\rm in}$ and $\langle Q_{\rm abs}\rangle_{\rm out}$ are
absorption efficiencies averaged over the incident (UV/optical) and emitted (IR) spectra. Using this radiative-equilibrium relation, the dust survival radius is
\[
R_{\rm sub}\approx
4.6\times10^{16}
\left(\frac{L_{\rm pk}}{10^{43.8}\ {\rm erg\,s^{-1}}}\right)^{1/2}
\left(\frac{T_{\rm sub}}{1800\ {\rm K}}\right)^{-2}
\ {\rm cm}.
\]
For the extinction-corrected peak luminosity, $\log(L_{\rm pk}/{\rm erg\,s^{-1}})\simeq43.8$, this gives
$R_{\rm sub}\approx4.6\times10^{16}$~cm $\approx3100$~AU, implying a distinct outer dusty component in addition to the compact inner CSM responsible for the rapid interaction.

\section{Discussion}
\label{sec:discussion}
\subsection{Dust origin scenario}

Here we explore the possible origin of the hot dust responsible for the near-infrared emission in SN~2018erx. Our inferred $T_{\rm d}\sim 1500$--$2000$~K is comparable to temperatures reported for hot dust components in other interacting SESNe \citep{Gan2021,Yamanaka2025}.
In SN~2006jc, the emergence of a red/NIR excess and contemporaneous spectroscopic changes were interpreted as rapid dust formation in the post-shock region, 
establishing that dust can condense on timescales of only weeks in a dense He-rich CSM interaction environment \citep{SmithFoley2008}. 
More recently, NIR and optical monitoring of the transitional Type~Ibn/Icn SN~2023xgo revealed a prominent NIR excess persisting from $\sim$15--100~d \citep{Yamanaka2025}, 
with SED modelling favouring carbon dust at $T_{\rm d}\approx 1600\pm100$~K and an inferred dust mass of $\sim 10^{-4}\,M_\odot$. 
That work argues that the NIR excess is best explained as an IR echo from pre-existing circumstellar carbon dust located outside the shocked CDS.
In the Type~Ibn SN OGLE-2012-SN-006, early optical--NIR SED modelling similarly favours a pre-existing dust shell, with dust temperatures of $\sim$1200--1300~K and substantially larger inferred dust masses ($\sim 5\times10^{-4}$--$2\times10^{-3}\,M_\odot$) \citep{Gan2021}. 

Thus, the hot dust mass we infer for SN~2018erx, $M_{\rm d}\sim10^{-6}$--$10^{-5}\,M_\odot$, lies toward the low end of the distribution, broadly comparable to the small dust masses or upper limits reported for several SNe~Ibn, and well below the $\sim10^{-4}$--$10^{-3}\,M_\odot$ values inferred for SN~2023xgo and OGLE-2012-SN-006. We caution that our two-band estimate constrains only the hot dust component emitting in $J$ and $K$; any cooler dust peaking at longer wavelengths would increase the total dust mass. Mid-IR photometry is therefore required to robustly constrain the full dust mass budget. If a fraction
$f_{\rm d}\sim1\%$ of the expelled material condensed into dust, the inferred
$M_{\rm d}\sim6\times10^{-7}$--$2\times10^{-5}\,M_\odot$ implies that the total
mass expelled in the dusty circumstellar layer was of order
$6\times10^{-5}$--$2\times10^{-3}\,M_\odot$. 

If the NIR excess is powered by a pre-existing IR echo, we estimated earlier that the dust responsible for the observed hot component must lie at radii of at least a few $\times10^{16}$~cm unless it is substantially shielded by optically thick gas or confined to dense clumps. This radius corresponds to a light-travel time of $R_{\rm sub}/c\sim10$--20~d and an echo duration of order $2R_{\rm sub}/c\sim20$--40~d for a thin shell, naturally consistent with the observations for SN~2018erx.

This in turn places a lower limit on when the dusty circumstellar layer was expelled prior to explosion. For a characteristic outflow velocity $v_{\rm w}$,
\begin{equation}
t_{\rm ej}\gtrsim \frac{R_{\rm sub}}{v_{\rm w}}
\approx 15
\left(\frac{R_{\rm sub}}{4.6\times10^{16}\ {\rm cm}}\right)
\left(\frac{v_{\rm w}}{10^3\ {\rm km\,s^{-1}}}\right)^{-1}
\ {\rm yr}.
\end{equation}
Thus, for WR-like outflow speeds ($v_{\rm w}\sim1000~{\rm km\,s^{-1}}$), the dusty layer must have been launched at least $\sim$10--20~yr before core collapse, while for slower outflows ($v_{\rm w}\sim100~{\rm km\,s^{-1}}$) the corresponding timescale becomes $\sim$100--200~yr. These estimates support a picture in which SN~2018erx is embedded in a structured, multi-epoch CSM: a compact inner C-rich component on $\lesssim$ tens-of-AU scales powers the rapid interaction, while an additional dusty component at much larger radii produces the large effective attenuation and the NIR excess. The fact that the optical lightcurve and spectra are already strongly reddened at the earliest observed epochs, together with line-profile asymmetries consistent with preferential attenuation of the receding side, strongly favors pre-existing circumstellar dust over dust formed only after explosion.


\subsection{Progenitor and evolutionary pathway}
\label{sec:progenitordiscussion}

SN~2018erx is one of the faintest members of the SN~Ibn/Icn sub-class. It rises and declines quickly with rise time and decline time of 2.1~d and 3.1~d respectively. Any viable progenitor scenario must account simultaneously for (i) the signatures of interaction with H/He-poor and likely C-rich material, and (ii) the small characteristic mass and compact spatial scale implied by our interaction fit. In particular, the \texttt{redback} CSM-interaction posteriors for SN~2018erx favor $M_{\rm ej}=0.11^{+0.03}_{-0.01}~M_\odot$ and $M_{\rm CSM}=0.30^{+0.01}_{-0.01}~M_\odot$. These values place SN~2018erx in a regime where the CSM mass is comparable to the ejecta mass, such that strong deceleration and efficient conversion of kinetic energy into radiation are expected. In addition, the rapidly fading light curve constrains the radioactive yield to be very small, with an upper limit of $M_{\rm Ni}\lesssim5.4\times10^{-3}\,M_\odot$ placing SN~2018erx at the low end of both the ejecta-mass and $^{56}$Ni-mass distributions among other SESNe \citep{Taddia2018, Das2023a, Das2023b, Das2024, Pellegrino2022b} (Figure~\ref{fig:mnimej}).

Below, we consider three broad progenitor and explosion pathways that could plausibly reproduce the observed and inferred properties of SN~2018erx: (1) an interaction-boosted ultra-stripped core-collapse supernova, (2) the core collapse of a more massive WC/WO Wolf--Rayet star, potentially involving fallback, and (3) a merger-powered transient involving a compact object and a stripped massive star.

\subsubsection{Ultra-stripped SN of a low-mass He-star}

SN~2018erx is consistent with a rapidly evolving, H-poor SN powered by the core-collapse of an ultra-stripped progenitor embedded in dense, nearby C-rich CSM. In the standard ultra-stripped channel, binary interaction removes almost the entire H-rich envelope and subsequently erodes much of the remaining He-rich layers during late Roche-lobe overflow (Case~BB/BC mass transfer) onto a compact companion, leaving behind a low-mass pre-SN core and therefore a very small ejecta mass at explosion \citep[e.g.,][]{Tauris2013,Tauris2015}. 

Despite allowing a very broad prior range ($0.1$--$15~M_\odot$), the fit for SN~2018erx favors a low ejecta mass,
$M_{\rm ej}=0.11^{+0.03}_{-0.01}~M_\odot$ (Table~\ref{tab:csm_posterior}). This places SN~2018erx firmly in the same ejecta-mass regime as well-studied ultra-stripped candidates such as iPTF\,14gqr ($M_{\rm ej}\sim 0.2~M_\odot$; \citealt{De2018c}), SN\,2019dge ($M_{\rm ej}\sim 0.4~M_\odot$; \citealt{Yao2020}), and SN\,2023zaw ($M_{\rm ej}\sim 0.5~M_\odot$; \citealt{Das2024}), and far below the typical $M_{\rm ej}\sim 1$--$5~M_\odot$ inferred for most SNe~Ib/c \citep[e.g.,][]{Taddia2018}. Such a small ejecta mass strongly suggests a progenitor that was stripped efficiently by a close companion rather than a single massive WR star retaining a more substantial CO core at collapse. Such a low value of ejecta mass has been seen previously for SN~2023xgo (SN Ibn/Icn; $M_{\rm ej}\sim 0.12~M_\odot$). While interaction complicates a purely radioactive interpretation, the low inferred $M_{\rm ej}$ itself is difficult to reconcile with a standard high-mass single-star channel.


At the same time, the fit favors a substantial circumstellar mass, $M_{\rm CSM}=0.30^{+0.01}_{-0.01}~M_\odot$ (Table~\ref{tab:csm_posterior}). This is few times the ejecta mass, with $M_{\rm CSM}/M_{\rm ej}\sim2.7$. Such a mass ratio is a qualitatively important regime for interaction: once the swept-up CSM mass approaches the ejecta mass, strong deceleration is expected and a significant fraction of the kinetic energy can be converted into radiation. In that limit, the emergent luminosity is governed primarily by shock interaction rather than radioactive heating, even if the underlying explosion is intrinsically faint. The parameter combination inferred for SN~2018erx therefore naturally places it in an interaction-dominated regime.

This configuration closely resembles the scenario explored by \citet{Moriya2025}, who demonstrate that an ultra-stripped explosion embedded in sufficiently massive nearby CSM can appear observationally as a luminous, H-poor interaction-powered transient. In their models, a pre-SN mass-ejection episode triggered by violent late burning produces CSM whose mass can rival or exceed that of the ejecta. The underlying mechanism is related to degenerate silicon flashes expected near the low-mass boundary for iron-core collapse. In mass-losing helium-star models, \citet{Woosley2019} show that helium cores of $M_{\rm He}\simeq2.5$--$3.2~M_\odot$ can undergo off-center silicon ignition under partially degenerate conditions, launching shocks that eject a substantial fraction of the remaining envelope. The resulting shell masses span $\sim0.02$--$0.7~M_\odot$, with characteristic ejection speeds of order $10^2$--$10^3~{\rm km~s^{-1}}$ and delays of tens of days between the flash and final collapse. These parameters naturally place dense material at radii $R\sim10^{14}$--$10^{15}$\,cm by the time of explosion. The best-fit $M_{\rm CSM}\approx0.3~M_\odot$ for SN~2018erx lies comfortably within this predicted shell-mass range, and the inferred compact CSM configuration is consistent with a mass-ejection episode occurring only months to years before core collapse. While the data do not uniquely require a silicon-flash origin, the overall scale of the inferred CSM mass and radius is at least compatible with this picture.

The fitted CSM density scale and inner radius further support a compact, recently ejected structure. The posterior medians, $\rho_0=(9.56^{+2.58}_{-1.12})\times10^{-13}~{\rm g~cm^{-3}}$ and $r_0=0.66^{+1.60}_{-0.47}~{\rm AU}$, set the scale for an optically thick, relatively compact CSM beginning just outside the progenitor. Approximating the CSM as a dense shell with mass $M_{\rm CSM}\sim0.3~M_\odot$ and characteristic density $\rho_0$ implies an outer radius of order a few $\times10^{14}$\,cm (tens of AU). This is far smaller than the $10^{16}$--$10^{17}$\,cm scales typical of long-lived interaction in SNe~IIn, yet entirely consistent with the confined, nearby CSM inferred for fast H-poor interaction events. For outflow speeds of $\sim100$--$1000~{\rm km~s^{-1}}$, such radii correspond to ejection occurring months to a few years before collapse, naturally pointing to very late-stage binary evolution. This aligns well with the calculations of \citet{Wu2022}, who model stripped helium stars through advanced burning and find that helium cores of $\simeq2.5$--$3~M_\odot$ can re-expand during O/Ne burning, triggering extreme late-stage mass transfer in close binaries. They obtain mass-transfer rates $\dot{M}\gtrsim10^{-2}~M_\odot~{\rm yr^{-1}}$ beginning weeks to decades before collapse, and show that either stable mass loss from the system or unstable transfer leading to a common-envelope ejection can produce CSM masses $\sim10^{-2}$--$1~M_\odot$ at radii $\sim10^{13}$--$10^{16}$\,cm. The posterior medians for SN~2018erx ($M_{\rm CSM}\simeq0.3~M_\odot$ and characteristic radii $\sim10^{14}$--$10^{15}$\,cm) fall well within these predicted ranges (Figure \ref{fig:csmmodels}). The mass-loss rates obtained from spectroscopy are also well within this range (Section~\ref{sec:mdot_inner_csm}). The spectroscopically derived $\dot{M}$ (Section~\ref{sec:mdot_inner_csm}) is consistent with ejection occurring months to a few years before collapse. The dust shell, sitting at larger radii, likely traces an earlier, distinct mass-loss episode -- possibly a prior phase of stable Roche-lobe overflow or a partial common-envelope interaction -- predating the terminal stripping event by decades to centuries. The consistency in mass scale and spatial extent suggests that the dense CSM around SN~2018erx could plausibly arise from the same late-stage binary processes that produce ultra-stripped explosions.

\subsubsection{Core collapse of a massive WC/WO Wolf--Rayet star}

A WC/WO progenitor remains attractive for SN~2018erx because it naturally explains the carbon-rich circumstellar environment implied by the strong intermediate-width \ion{C}{2} emission and, plausibly, the carbonaceous dust indicated by the NIR excess. In particular, a WC/WO-like surface composition provides a straightforward route to H/He-poor, C/O-rich CSM, unlike lower-mass stripped progenitors for which the origin of carbon-rich outflows is less direct \citep[e.g.,][]{Galyam2022,PerleyIcn,Pellegrino2022b}. The outer dusty layer inferred from the NIR echo in SN~2018erx is also not, by itself, problematic for such a progenitor: for WR-like outflow speeds, the dust-survival radius implies ejection on a timescale of order $\sim$10--20 yr before core collapse, while the hot dust mass implies only a modest total dusty outflow mass.

The difficulty lies instead in the inner interaction region. Our modeling requires $M_{\rm CSM}\approx0.3\,M_\odot$, $R_0\approx0.7$~AU, $R_{\rm out}\sim40$~AU, and a nearly constant-density profile ($\eta\approx0$), i.e.\ a compact shell rather than a steady $r^{-2}$ wind. Placing $\sim0.3\,M_\odot$ within $\sim6\times10^{14}$~cm implies characteristic mass-loss rates of order $\dot{M}\sim0.1$--$1\,M_\odot\,{\rm yr^{-1}}$ for $v_w\sim10^2$--$10^3~{\rm km\,s^{-1}}$, far above canonical steady WR wind values \citep[e.g.,][]{Crowther2007}. Likewise, the shell-like density structure inferred here is unlike the wind-like CSM expected from a long-lived WR outflow. In this sense, a standard single-star WC/WO wind is difficult to reconcile with SN~2018erx; if the progenitor was WR-like at collapse, the data require an unusually violent pre-SN ejection episode, strong binary mediation, or both.

A second tension is the explosion scale itself. The inferred ejecta mass, $M_{\rm ej}\approx0.11\,M_\odot$, is far below that expected from a normal successful explosion of a massive WC/WO star, which would typically retain a substantially larger CO core at collapse. Thus, while a WC/WO-like surface composition remains an appealing explanation for the carbon-rich CSM, the combined requirements of a compact shell-like CSM and a very low ejecta mass argue against a simple steady-wind WR channel. If SN~2018erx did originate from a massive WR progenitor, then fallback or some other mechanism for suppressing the effective ejecta mass is likely required (see next subsection).

\subsubsection{A partially successful (fallback) explosion and jet-driven outflow}

In fallback SN scenarios, the explosion energy is insufficient to unbind the entire star; the inner layers fall back onto the compact remnant, reducing both the effective ejecta mass and the $^{56}$Ni yield \citep[e.g.,][]{Woosley1993,Moriya2010}. A jet launched during or after collapse \citep{Khokhlov1999,MacFadyen1999,MacFadyen2001} can still imprint broad features even if the bulk explosion is weak, linking some SNe~Icn/Ibn to peculiar Ic-BL-like events with low $^{56}$Ni yields \citep[e.g.,][]{Fraser2021,perley2022,Pellegrino2022b}. In the context of SN~2018erx, fallback offers a coherent explanation for the combination of low ejecta mass, weak radioactive output, and low expansion velocities, while remaining compatible with a WC/WO progenitor and hence a carbon-rich circumstellar environment.

For SN~2018erx, the inferred $M_{\rm ej}\approx0.11~M_\odot$ and $M_{\rm Ni}\lesssim5.4\times10^{-3}~M_\odot$ lie firmly in the regime expected for strong fallback, where a significant fraction of the initially ejected material fails to escape. The observed photospheric velocity of $\sim5200~\mathrm{km\,s^{-1}}$ is also consistent with a low-energy explosion in which the surviving ejecta expand more slowly than in typical stripped-envelope SNe. 

However, the compact, shell-like CSM ($M_{\rm CSM}\approx0.3~M_\odot$, $R_{\rm out}\sim40$~AU, $\eta\approx0$) still requires a brief and intense pre-SN mass ejection episode, as steady WR winds cannot deposit $\sim0.3~M_\odot$ at such small radii. Fallback therefore constrains the explosion outcome, but does not by itself explain the origin of the dense nearby CSM, and is most naturally combined with a binary-mediated mass-loss channel.

\subsubsection{Merger-powered fast transients involving a compact object}

A merger-powered scenario is appealing for SN~2018erx primarily because it can naturally produce the structured, asymmetric, and multi-epoch circumstellar environment implied by the data. The event appears to require both a dense inner shell on sub-AU to tens-of-AU scales, which powers the rapid interaction, and a distinct outer dusty layer at radii of at least a few $\times10^{16}$~cm ($M_{\rm d}\sim10^{-6}$--$10^{-5}~M_\odot$), which accounts for the strong effective attenuation ($A_V\simeq3.3$~mag) and the NIR echo. In particular, delayed mergers between a stripped WR star and a compact object (neutron star or black hole) provide a physically motivated channel, in which the WR star is tidally disrupted and accreted, driving a disk-wind outflow and shaping both the inner and outer CSM \citep{Metzger2022}. 

The merger picture is also attractive because it does not require the inner CSM to resemble a steady wind. In SN~2018erx the inferred shell-like density profile ($\eta\approx0.05$), compact inner radius ($R_0\approx0.7$~AU), and relatively high CSM density ($\rho_0\approx9.6\times10^{-13}~{\rm g~cm^{-3}}$) are more suggestive of a confined mass ejection than of a long-lived outflow. The implied characteristic mass-loss rate, $\dot{M}\sim10^{-3}~M_\odot~{\rm yr^{-1}}$ (Section~\ref{sec:mdot_inner_csm}), is orders 
of magnitude above canonical WR wind values, yet consistent with rates expected during 
a common-envelope or dynamical mass-transfer episode. In this sense, merger-related ejection or common-envelope clearing may better match the geometry of the inferred CSM than a canonical WR wind. If the stripped star involved in the merger were WC/WO-like, the carbon-rich line spectrum and carbonaceous dust would also follow naturally.

The main limitation is that current merger models are not yet tied closely enough to the specific parameter combination inferred for SN~2018erx. It remains unclear whether they generically produce an effective ejecta mass as low as $M_{\rm ej}\sim0.11\,M_\odot$ together with $M_{\rm CSM}\sim0.3\,M_\odot$ at radii of only tens of AU, while also reproducing Type~Icn-like intermediate-width \ion{C}{2} emission. We also lack independent observational evidence for merger or engine signatures such as luminous X-ray or radio emission, long-lived central-engine activity, or extreme late-time asymmetry, although the available follow-up is not deep enough to exclude them. We therefore regard merger-powered models as plausible, particularly for explaining the CSM geometry and the presence of a distinct dusty outer layer, but not yet uniquely required by the present data.

\begin{figure*}
    \centering
    \includegraphics[width=0.7\textwidth]{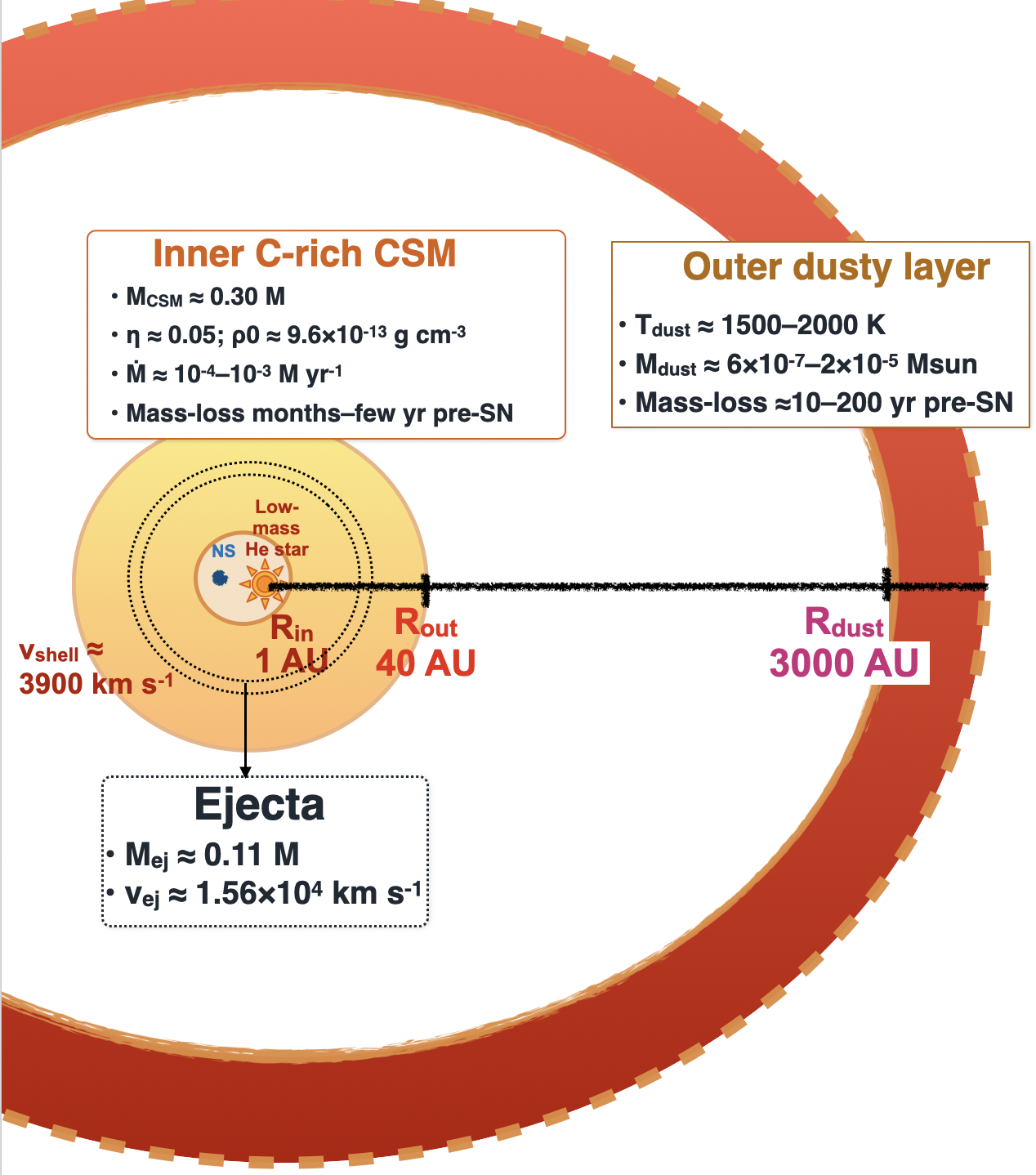}
    \caption{Schematic illustration of the CSM environment inferred for SN~2018erx from the lightcurve and spectral modeling. The figure summarizes the compact C-rich inner CSM responsible for the rapid interaction, together with the more extended dusty component inferred from the strong extinction and late-time near-infrared excess.}
    \label{fig:cartoon}
\end{figure*}

\section{Conclusion}
\label{sec:conclusion}
We have presented optical and near-infrared observations of SN~2018erx (ZTF18abkmbpy), a fast-evolving and unusually red interacting SESN discovered by ZTF. Photometrically, SN~2018erx rises to peak in $\sim 2$~d and declines at $0.16\pm0.01$~mag~d$^{-1}$ over the first 10~d, placing it among the most rapidly evolving hydrogen-poor SNe and overlapping with the parameter space of rapidly evolving stripped-envelope transients. Spectroscopically, it exhibits prominent, relatively broad \ion{C}{2} emission with characteristic widths of $\sim3800$~km~s$^{-1}$, consistent with interaction between the SN ejecta and a carbon-rich CSM and favoring a Type~Icn classification. Semi-analytical CSM-interaction modeling indicates a compact, shell-like CSM with $M_{\rm CSM}\approx0.3\,M_\odot$ interacting with low-mass ejecta of $M_{\rm ej}\approx0.11\,M_\odot$ at velocities of $v_{\rm ej}\approx1.5\times10^{4}$~km~s$^{-1}$, while the late-time luminosity constrains the $^{56}$Ni mass to $\lesssim(3$–$5)\times10^{-3}\,M_\odot$. The estimated near-constant density profile ($\eta\approx0.05$) and high inner density ($\rho_0\approx9.6\times10^{-13}$~g~cm$^{-3}$) disfavor a steady WR wind and instead 
point to a brief, intense mass-ejection episode in the final years before collapse, most naturally explained by late-stage binary interaction. The strong continuum reddening ($A_V\simeq3.3$~mag) and near-infrared excess at $+27$~d, corresponding to a hot dust component at $T_{\rm d}\sim1500$--$2000$~K with inferred mass $M_{\rm d}\sim10^{-6}$--$10^{-5}\,M_\odot$, point to local circumstellar dust. The 
dust sublimation radius ($R_{\rm sub}\approx4.6\times10^{16}$~cm) implies that this outer dusty layer was ejected $\sim10$--$200$~yr before core collapse, representing an 
earlier mass-loss episode distinct from that of the compact inner CSM. Together, the rapid evolution, low ejecta mass, modest radioactive yield, and evidence for compact CSM interaction place SN~2018erx in a sparsely populated region of the interacting SESN phase space.

The inferred properties are most naturally consistent with an interaction-boosted ultra-stripped core-collapse explosion in a close binary system. In particular, the very low ejecta mass ($M_{\rm ej}\sim0.11~M_\odot$), very small radioactive yield ($M_{\rm Ni}\lesssim(3$--$5)\times10^{-3}~M_\odot$), rapid photometric evolution, and compact dense CSM with a mass comparable to that of the ejecta are all consistent with a progenitor stripped to a low-mass core via binary interaction, followed by a final episode of intense mass loss through Case~BC evolution or a silicon-flash-driven ejection shortly before collapse. At the same time, the carbon-rich circumstellar environment is not uniquely diagnostic of this channel. A WC/WO progenitor provides a direct route to C-rich gas and dust, but a steady Wolf--Rayet wind struggles to produce the inferred combination of low $^{56}$Ni mass, compact shell-like CSM with an approximately constant density profile, and such a small ejecta mass without invoking an additional eruptive mass-loss episode and/or strong fallback. A fallback-modified WR explosion can account for the low effective ejecta mass and suppressed $^{56}$Ni production, yet it does not by itself explain the origin of the dense nearby shell required by the interaction. Merger- or compact-object-powered scenarios remain plausible because they can generate a structured, multi-epoch circumstellar environment, including both the compact inner shell and the older dusty outer material. These scenarios also naturally accommodate the high inferred mass-loss rates and shell-like geometry. However, current models do not yet demonstrate that they reproduce the full combination of low ejecta mass, low nickel yield, compact CSM, and Type~Icn-like spectroscopic properties observed in SN~2018erx. Overall, the ultra-stripped interpretation provides the most self-consistent explanation of the full parameter set, while WR/fallback and merger-driven channels remain viable alternatives that likely require more detailed modelling.

Looking ahead, distinguishing between an interaction-boosted ultra-stripped explosion, a fallback-modified Wolf–Rayet collapse forming a black hole, or a merger/engine-driven scenario will require more detailed modelling and substantially deeper late-time observations. In particular, very deep optical and near-infrared tail photometry is essential to measure or tightly constrain any $^{56}$Co-powered decay. This will allow us to separate a zero nickel yield from a merely low $^{56}$Ni mass. Late-time spectroscopy is equally important, as nebular line diagnostics can directly probe the inner ejecta composition, progenitor mass, and the persistence or shutdown of circumstellar interaction.

The detection of local circumstellar dust in SN~2018erx raises a broader question about the role of dust in the interacting SESN population. While dust signatures have been identified in individual Type~Ibn and Icn events (e.g., SN~2006jc, OGLE-2012-SN-006, SN~2023xgo), they are not seen in all cases. This likely reflects a combination of factors. The dust may be distributed in a clumpy or aspherical configuration, so that it does not lie along every line of sight. In other events, the total dust mass may simply be too low to produce a detectable signature at optical wavelengths. Observational biases also play an important role: heavily dust-reddened events like SN~2018erx appear intrinsically faint in the optical before extinction correction, making them easier to miss or misclassify in optical surveys. Multi-epoch near- and mid-infrared monitoring will be crucial to fully characterize this dust component. Type~Icn SNe, with their dense and carbon-rich C/O-dominated circumstellar environments, may represent especially favorable conditions for efficient dust condensation and growth, making them promising targets for JWST follow-up. Robust modeling will connect dust formation to the progenitor’s final mass-loss history and, more broadly, these carbon-rich environments provide a direct probe of whether core-collapse SNe can significantly contribute to the cosmic dust budget.

Finally, SN~2018erx adds to the growing evidence that stripped-envelope explosions with very low $^{56}$Ni masses may be underrepresented in current samples. Previous studies have suggested a discrepancy between nickel yields in stripped-envelope SNe and Type~II SNe \citep{Anderson2019}, possibly driven by observational bias. SN~2018erx reinforces the emerging picture that low-nickel, low-ejecta-mass events are faint and evolve rapidly, and may only be discovered or classified when their peak luminosity is boosted by circumstellar interaction or shock cooling. Similarly, SN~2018erx shows that sufficiently high dust extinction can make a Type~Icn SN appear optically faint and unusually red, suggesting that dustier members of the class may be systematically missed in optical surveys. Future deep surveys such as LSST \citep{Ivezic08} and Roman, combined with high-cadence surveys such as ZTF, will be transformative in uncovering this hidden population.



\section{Data availability}
All photometry and spectroscopy data will be made publicly available on Zenodo and WISeREP \citep{Yaron:2012aa} after publication.

\section{Acknowldegements}

Based on observations obtained with the Samuel Oschin Telescope 48-inch and the 60-inch Telescope at the Palomar Observatory as part of the Zwicky Transient Facility project. ZTF is supported by the National Science Foundation under Grant No. AST-1440341 and a collaboration including Caltech, IPAC, the Weizmann Institute of Science, the Oskar Klein Center at Stockholm University, the University of Maryland, the University of Washington, Deutsches Elektronen-Synchrotron and Humboldt University, Los Alamos National Laboratories, the TANGO Consortium of Taiwan, the University of Wisconsin at Milwaukee, and Lawrence Berkeley National Laboratories. Operations are conducted by COO, IPAC, and UW.

SED Machine is based upon work supported by the National Science Foundation under Grant No. 1106171

The ZTF forced-photometry service was funded under the Heising-Simons Foundation grant 12540303 (PI: Graham).

The Gordon and Betty Moore Foundation, through both the Data-Driven Investigator Program and a dedicated grant, provided critical funding for SkyPortal.


M.W.C. acknowledges support from the National Science Foundation with grant numbers PHY-2117997, PHY-2308862 and PHY-2409481.

D. T. is supported by Harvard University through the Institute for Theory and Computation Fellowship.

The Gordon and Betty Moore Foundation, through both the Data-Driven Investigator Program and a dedicated grant, provided critical funding for SkyPortal .

This research has made use of the NASA/IPAC Extragalactic Database (NED), which is funded by the National Aeronautics and Space Administration and operated by the California Institute of Technology.

The Liverpool Telescope is operated on the island of La Palma by Liverpool John Moores University in the Spanish Observatorio del Roque de los Muchachos of the Instituto de Astrofisica de Canarias with financial support from the UK Science and Technology Facilities Council.

The W. M. Keck Observatory is operated as a scientific partnership among the California Institute of Technology, the University of California and the National Aeronautics and Space Administration. The Observatory was made possible by the generous financial support of the W. M. Keck Foundation. The authors wish to recognize and acknowledge the very significant cultural role and reverence that the summit of Maunakea has always had within the indigenous Hawaiian community.  We are most fortunate to have the opportunity to conduct observations from this mountain.

\textit{Software:} Global Relay of Observatories Watching Transients Happen Marshal \citep[GROWTH;][]{Kasliwal2019} and the Fritz SkyPortal Marshal  \citep[][]{Duev2019, skyportal, Coughlin2023}. Astropy \citep{Astropy-Collaboration13}, Matplotlib \citep{Hunter07}

\FloatBarrier
\bibliography{main}
\bibliographystyle{aasjournal}

\appendix

\setcounter{table}{0}
\setcounter{figure}{0}

\renewcommand{\thetable}{A\arabic{table}}
\renewcommand{\thefigure}{A\arabic{figure}}

\begin{table*}
\centering
\caption{Photometry after MW extinction correction.}
\label{tab:all_phot}
\setlength{\tabcolsep}{5pt}
\renewcommand{\arraystretch}{1.05}
\begin{tabular}{r r l l r r r l l r}
\toprule
Phase (d) & MJD & Instrument & Filter & AB mag & Phase (d) & MJD & Instrument & Filter & AB mag \\
\midrule
-2.15 & 58331.17716 & P48+ZTF & $r$ & 20.32 $\pm$ 0.18 & 0.00 & 58333.32784 & P48+ZTF & $r$ & 18.54 $\pm$ 0.04 \\
-2.15 & 58331.17716 & P48+ZTF & $r$ & 20.25 $\pm$ 0.11 & 2.89 & 58336.21634 & P48+ZTF & $r$ & 18.74 $\pm$ 0.06 \\
-2.15 & 58331.17716 & P48+ZTF & $r$ & 20.25 $\pm$ 0.11 & 2.89 & 58336.21634 & P48+ZTF & $r$ & 18.67 $\pm$ 0.03 \\
-2.13 & 58331.20277 & P48+ZTF & $r$ & 20.40 $\pm$ 0.24 & 2.89 & 58336.21634 & P48+ZTF & $r$ & 18.67 $\pm$ 0.03 \\
-2.13 & 58331.20277 & P48+ZTF & $r$ & 20.30 $\pm$ 0.14 & 2.89 & 58336.22250 & P60 & $r$ & 18.89 $\pm$ 0.03 \\
-2.13 & 58331.20277 & P48+ZTF & $r$ & 20.30 $\pm$ 0.14 & 2.90 & 58336.22530 & P60 & $g$ & 19.78 $\pm$ 0.05 \\
-2.11 & 58331.21819 & P48+ZTF & $r$ & 20.25 $\pm$ 0.18 & 2.90 & 58336.22810 & P60 & $i$ & 18.40 $\pm$ 0.03 \\
-2.11 & 58331.21819 & P48+ZTF & $r$ & 20.22 $\pm$ 0.12 & 2.91 & 58336.24016 & P48+ZTF & $g$ & 19.90 $\pm$ 0.10 \\
-2.11 & 58331.21819 & P48+ZTF & $r$ & 20.22 $\pm$ 0.12 & 4.83 & 58338.16157 & P48+ZTF & $i$ & 18.62 $\pm$ 0.06 \\
-2.00 & 58331.32586 & P48+ZTF & $i$ & 19.47 $\pm$ 0.13 & 6.85 & 58340.17862 & P48+ZTF & $r$ & 19.44 $\pm$ 0.16 \\
-1.94 & 58331.38350 & P60 & $r$ & 20.02 $\pm$ 0.10 & 6.85 & 58340.17862 & P48+ZTF & $r$ & 19.39 $\pm$ 0.08 \\
-1.94 & 58331.38900 & P60 & $i$ & 19.42 $\pm$ 0.07 & 6.85 & 58340.17862 & P48+ZTF & $r$ & 19.39 $\pm$ 0.08 \\
-1.13 & 58332.20088 & P48+ZTF & $g$ & 20.00 $\pm$ 0.11 & 6.92 & 58340.25020 & P60 & $r$ & 19.66 $\pm$ 0.05 \\
-1.13 & 58332.20088 & P48+ZTF & $g$ & 19.95 $\pm$ 0.20 & 6.93 & 58340.25300 & P60 & $g$ & 20.57 $\pm$ 0.09 \\
-1.13 & 58332.20088 & P48+ZTF & $g$ & 20.00 $\pm$ 0.11 & 6.93 & 58340.25570 & P60 & $i$ & 19.14 $\pm$ 0.04 \\
-1.11 & 58332.22079 & P48+ZTF & $g$ & 19.99 $\pm$ 0.10 & 8.99 & 58342.31620 & P60 & $r$ & 20.17 $\pm$ 0.08 \\
-1.11 & 58332.22079 & P48+ZTF & $g$ & 19.93 $\pm$ 0.22 & 8.99 & 58342.31900 & P60 & $g$ & 20.97 $\pm$ 0.10 \\
-1.11 & 58332.22079 & P48+ZTF & $g$ & 19.99 $\pm$ 0.10 & 8.99 & 58342.32170 & P60 & $i$ & 19.71 $\pm$ 0.08 \\
-1.07 & 58332.26127 & P48+ZTF & $r$ & 19.02 $\pm$ 0.09 & 9.04 & 58342.36434 & P48+ZTF & $i$ & 19.61 $\pm$ 0.17 \\
-1.07 & 58332.26127 & P48+ZTF & $r$ & 18.91 $\pm$ 0.05 & 9.83 & 58343.15726 & P48+ZTF & $r$ & 20.18 $\pm$ 0.12 \\
-1.05 & 58332.28213 & P48+ZTF & $r$ & 18.86 $\pm$ 0.05 & 9.83 & 58343.15726 & P48+ZTF & $r$ & 20.18 $\pm$ 0.12 \\
-1.05 & 58332.28213 & P48+ZTF & $r$ & 18.86 $\pm$ 0.05 & 12.57 & 58345.89690 & LT+IOO & $g$ & 22.34 $\pm$ 0.20 \\
-1.03 & 58332.30164 & P48+ZTF & $r$ & 18.74 $\pm$ 0.12 & 12.57 & 58345.89890 & LT+IOO & $r$ & 21.72 $\pm$ 0.15 \\
-0.05 & 58333.27800 & P60 & $r$ & 18.72 $\pm$ 0.02 & 12.57 & 58345.90090 & LT+IOO & $i$ & 21.09 $\pm$ 0.10 \\
-0.05 & 58333.28090 & P60 & $g$ & 19.57 $\pm$ 0.03 & 12.85 & 58346.17784 & P48+ZTF & $r$ & 20.90 $\pm$ 0.20 \\
-0.04 & 58333.28360 & P60 & $i$ & 18.26 $\pm$ 0.03 & 28.94 & 58362.26390 & Keck & $K$ & 21.23 $\pm$ 0.03 \\
0.00 & 58333.32784 & P48+ZTF & $r$ & 18.65 $\pm$ 0.09 & 28.95 & 58362.27590 & Keck & $J$ & 22.64 $\pm$ 0.13 \\
0.00 & 58333.32784 & P48+ZTF & $r$ & 18.54 $\pm$ 0.04 &  &  &  &  &  \\
\bottomrule
\end{tabular}
\end{table*}

\begin{table}[t]
\centering
\caption{Spectroscopy log}
\begin{tabular}{lll}
\hline
Date & Phase (d) & Telescope/Instrument \\
\hline
2018-08-03 & -1.6 & P60/SEDM \\
2018-08-04 & -0.6 & P200/DBSP \\
2018-08-07 & +2.4 & Keck-I/LRIS \\
2018-08-12 & +7.4 & P200/DBSP \\
2018-08-21 & +16.4 & P200/DBSP \\
2018-09-02 & +29.4 & Keck-II/NIRES \\
2025-08-19 & +2572.7 & Keck-I/LRIS \\
\hline
\end{tabular}
\label{tab:spec_log}
\end{table}

\begin{table*}[ht]
\centering
\caption{Blackbody evolution of SN 2018erx before and after extinction correction. Blank entries did not pass the quality cut $\sigma_{L_{\rm bb}}/L_{\rm bb}<1$ because their best-fit luminosity uncertainties are too large.}
\label{tab:bbtable}
\begin{tabular}{ccccccc}
\hline
Phase & \multicolumn{3}{c}{Before correction ($A_V=0$)} & \multicolumn{3}{c}{After correction ($A_V=3.3$)} \\
(days) & $\log L_{\rm bb}$ & $T_{\rm bb}$ & $R_{\rm bb}$ & $\log L_{\rm bb}$ & $T_{\rm bb}$ & $R_{\rm bb}$ \\
 & (erg s$^{-1}$) & (K) & ($R_\odot$) & (erg s$^{-1}$) & (K) & ($R_\odot$) \\
\hline
$-3.3$ & $41.96^{+0.06}_{-0.06}$ & $3130^{+150}_{-140}$ & $52380^{+8650}_{-7650}$ & $42.72^{+0.12}_{-0.03}$ & $6840^{+2120}_{-730}$ & $26370^{+5530}_{-8720}$ \\
$-2.3$ & $42.27^{+0.03}_{-0.03}$ & $3900^{+100}_{-100}$ & $48180^{+4340}_{-3810}$ & $43.38^{+0.04}_{-0.03}$ & $9020^{+770}_{-610}$ & $32270^{+3560}_{-3550}$ \\
$-1.3$ & $42.28^{+0.01}_{-0.01}$ & $4340^{+60}_{-60}$ & $39350^{+1570}_{-1510}$ & $43.78^{+0.05}_{-0.05}$ & $14320^{+840}_{-790}$ & $20440^{+1240}_{-1140}$ \\
$1.7$ & $42.25^{+0.01}_{-0.01}$ & $4210^{+70}_{-70}$ & $40450^{+1850}_{-1750}$ & $43.67^{+0.05}_{-0.05}$ & $13200^{+840}_{-750}$ & $21030^{+1360}_{-1300}$ \\
$5.6$ & $41.94^{+0.02}_{-0.02}$ & $4180^{+120}_{-120}$ & $28640^{+2280}_{-2010}$ & $43.38^{+0.11}_{-0.09}$ & $13650^{+1830}_{-1390}$ & $14090^{+1640}_{-1620}$ \\
$7.8$ & $41.73^{+0.05}_{-0.04}$ & $4010^{+370}_{-330}$ & $24370^{+6160}_{-4700}$ & $\cdots$ & $\cdots$ & $\cdots$ \\
$11.4$ & $41.14^{+0.05}_{-0.04}$ & $4330^{+350}_{-330}$ & $10620^{+2430}_{-1860}$ & $\cdots$ & $\cdots$ & $\cdots$ \\
$27.7$ & $40.63^{+0.01}_{-0.01}$ & $1660^{+60}_{-70}$ & $40200^{+3770}_{-3020}$ & $40.81^{+0.02}_{-0.02}$ & $2020^{+100}_{-100}$ & $33330^{+3290}_{-2620}$ \\
\hline
\end{tabular}
\end{table*}

\begin{figure}
    \centering
    \includegraphics[width=8.5cm]{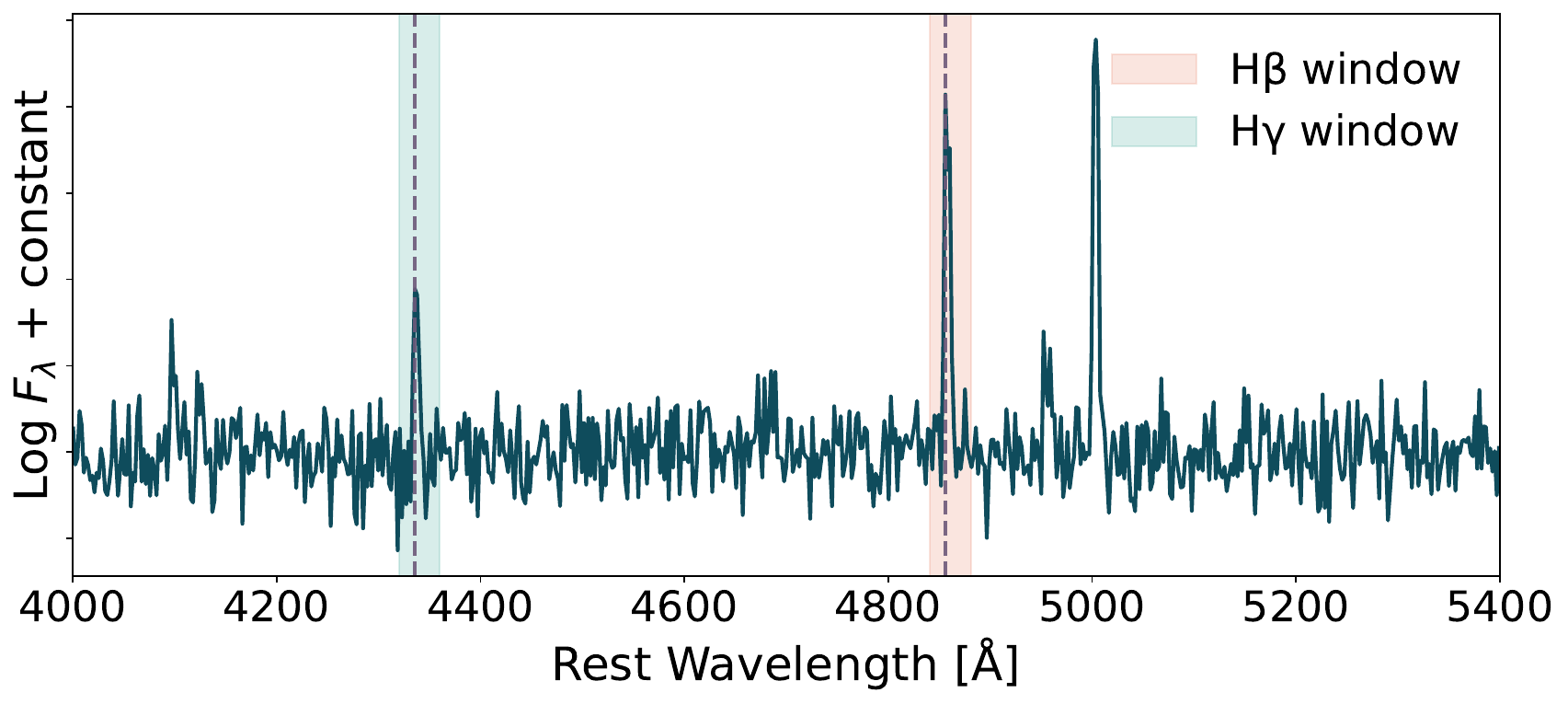}
    \caption{Measurement of the Balmer decrement in the host-galaxy spectrum at the position of SN~2018erx using a late-time Keck/LRIS observation. The continuum-subtracted host spectrum used to measure the line fluxes. The shaded regions mark the wavelength windows used to integrate the H$\beta$ and H$\gamma$ emission lines.}
    \label{fig:host}
\end{figure}

\begin{figure}
    \centering
    \includegraphics[width=\columnwidth]{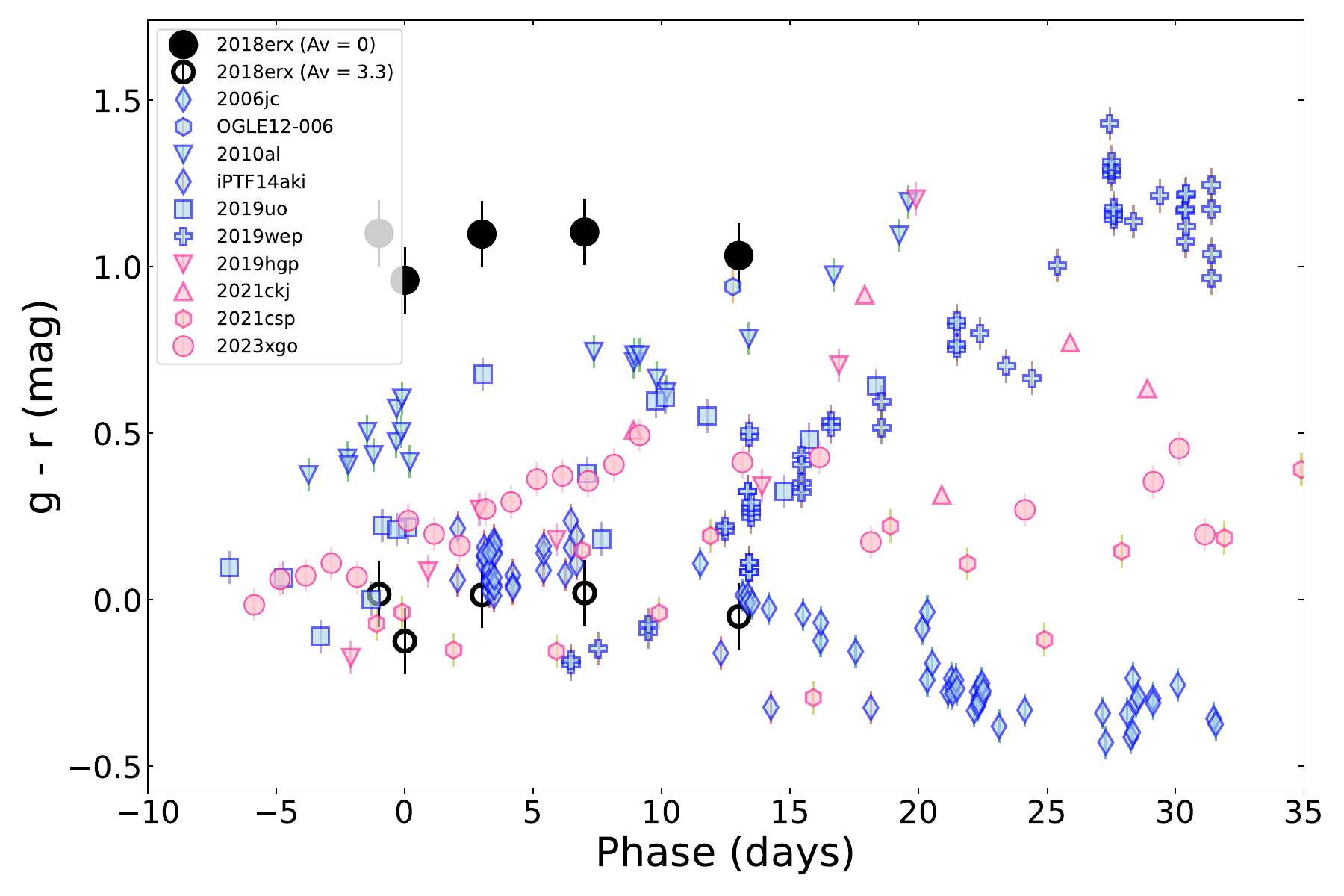}
    \caption{Evolution of the optical color $(g-r)$ for SN~2018erx compared with Type~Ibn (blue symbols) and Type~Icn (pink symbols) supernovae from the literature. 
Filled black circles show the observed colors of SN~2018erx assuming no host extinction ($A_V=0$), while open black circles show the extinction-corrected values adopting $A_V=3.3$~mag. 
Phases are given relative to $r$-band maximum. 
Compared to the comparison samples, SN~2018erx exhibits unusually red colors if no host extinction is applied, while the extinction-corrected colors are broadly consistent with the color evolution of other interacting SESNe.
}
    \label{fig:color}
\end{figure}

\begin{figure*}
    \centering
    \includegraphics[width=9cm]{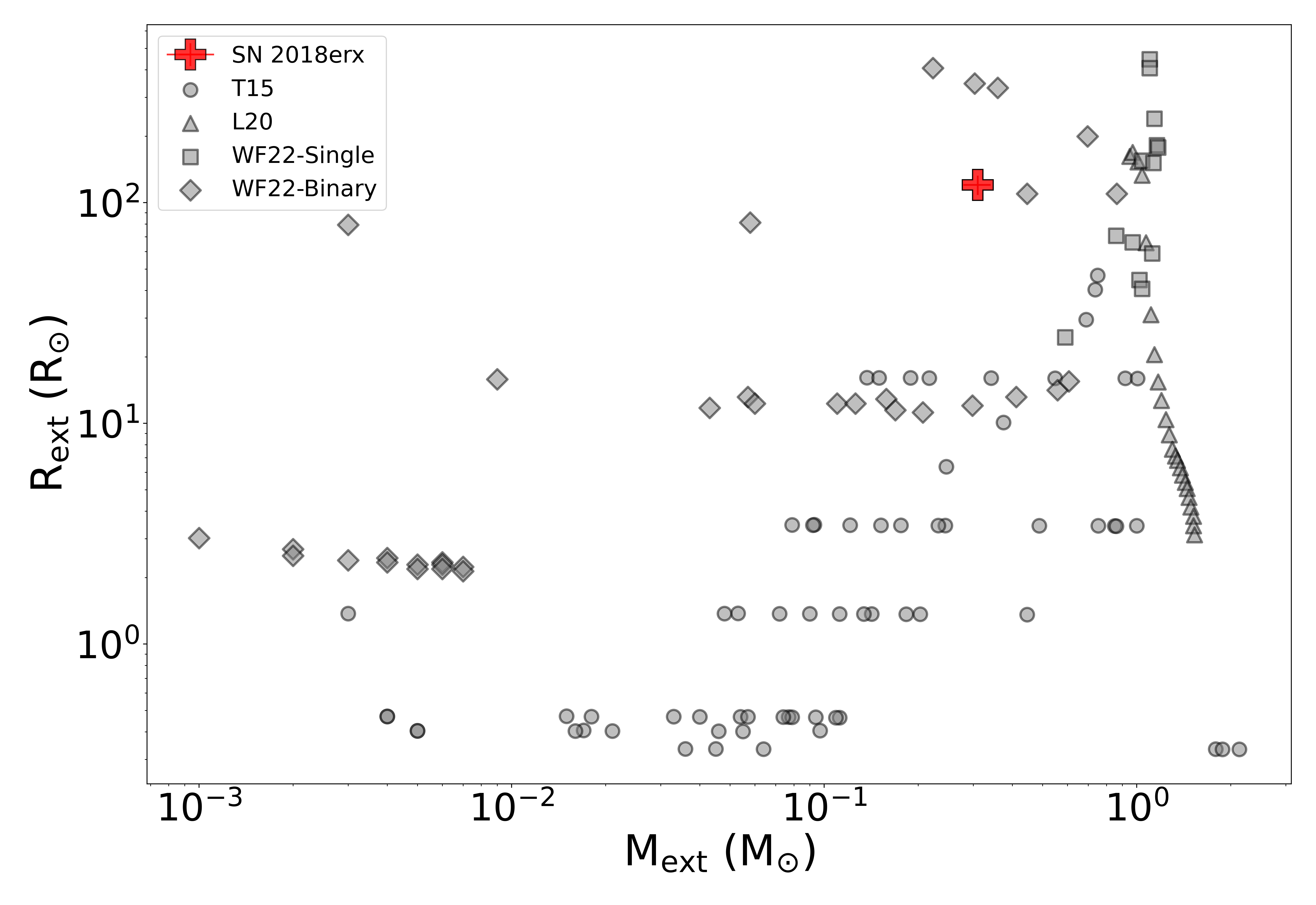}\includegraphics[width=9cm]{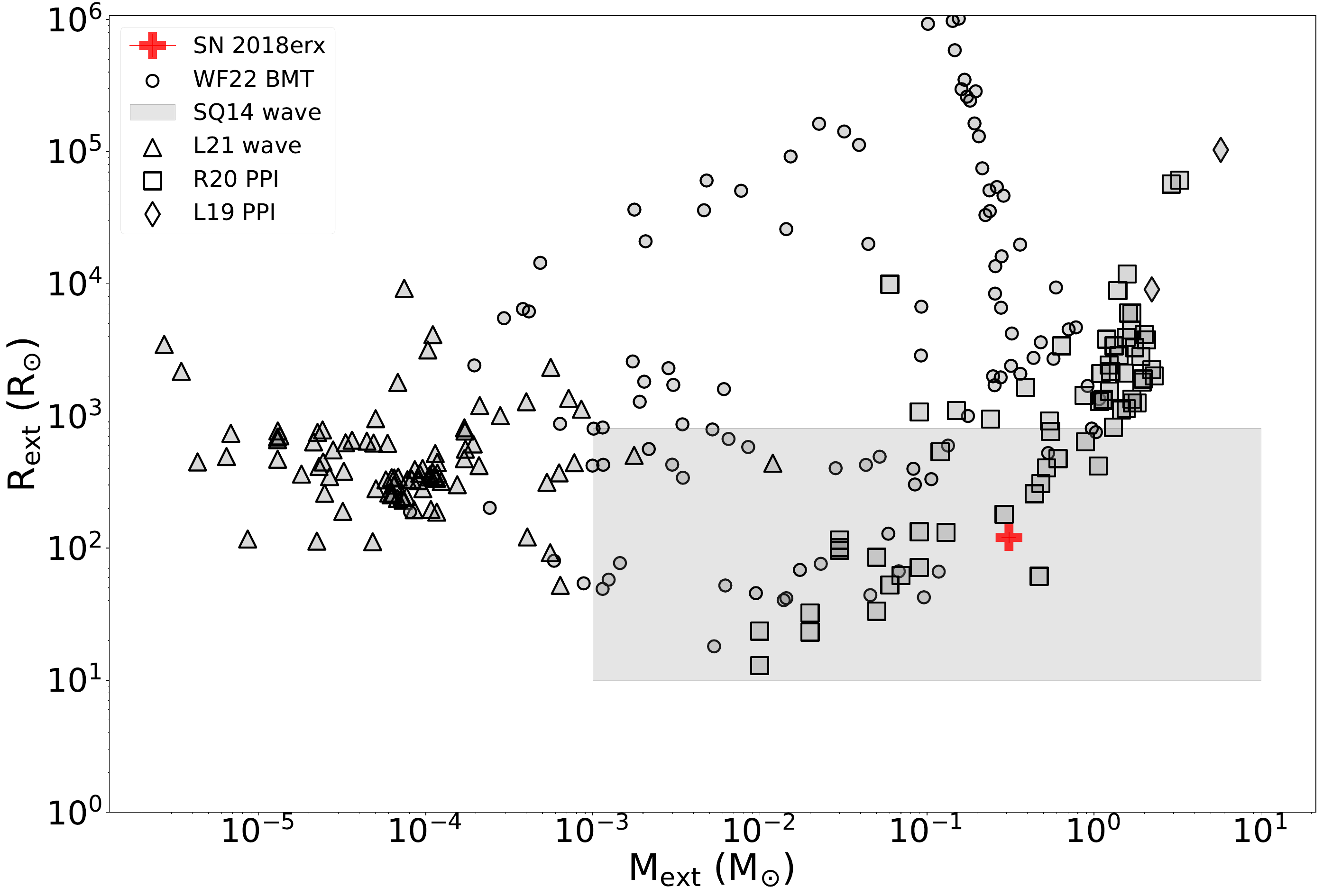}
    \caption{Comparison of the envelope properties of SN~2018erx with theoretical models for \textit{bound} stellar material $-$ binary and single star models from \citet{Wu2022b} (WF22), \citet{Laplace2020} (L20) and \citet{Tauris2015} (T15). \textit{Right:} Comparison of the the envelope properties of SN 2023zaw with theoretical models for \textit{unbound} stellar material $-$ late-time binary mass transfer \citep[BMT;][]{Wu2022b}, wave-driven mass loss \citep{Leung2021b, Shiode2014}, pulsation-pair instability driven  mass loss \citep[PPI;][]{Renzo2020, Leung2019}.}
    \label{fig:csmmodels}
\end{figure*}

\end{document}